\documentclass[fleqn]{mn2e}
\usepackage{epsfig,natbib_vw,times}
\usepackage{amssymb,amsmath}

\bibpunct[, ]{(}{)}{;}{a}{}{,}

\def \be {\begin{equation}}
\def \ee {\end{equation}}

\def\gsim{\mathrel{\lower0.6ex\hbox{$\buildrel {\textstyle >}
 \over {\scriptstyle \sim}$}}}
\def\lsim{\mathrel{\lower0.6ex\hbox{$\buildrel {\textstyle <}
 \over {\scriptstyle \sim}$}}}

\def\m@th{\mathsurround=0pt }
\def\eqalign#1{\null\,\vcenter{\openup1\jot \m@th
 \ialign{\strut\hfil$\displaystyle{##}$&$\displaystyle{{}##}$\hfil
 \crcr#1\crcr}}\,}

\def \hda {H$\delta_{\rm A}$}

\def \he {H$\epsilon$}

\def \ha {H$\alpha$}
\def \nii {[N~{\sc ii}]}
\def \sii {[S~{\sc ii}]}
\def \neiii {[Ne~{\sc iii}]}
\def \oiii {[O~{\sc iii}]}
\def \oii {[O~{\sc ii}]}

\def \mum {$\mu$m}

\title[A new method for classifying galaxy SEDs]
{A new method for classifying galaxy SEDs from
  multi-wavelength photometry} \author[V. Wild et al.]{ 
\parbox[t]{\textwidth}{\raggedright Vivienne
  Wild$^{1,2}$\thanks{vw8@st-andrews.ac.uk}, Omar Almaini$^{3}$, Michele Cirasuolo$^{4}$, Jim Dunlop$^{2}$, 
  Ross McLure$^{2}$, Rebecca Bowler$^{2}$, Joao Ferreira$^{2}$, Emma
  Bradshaw$^{3}$, Robert Chuter$^{3}$, Will Hartley$^{3,5}$
}
\vspace*{6pt}\\
$^1$School of Physics and Astronomy, University of St Andrews, North
Haugh, St Andrews, KY16 9SS, U.K. (SUPA\thanks{Scottish Universities
  Physics Alliance})\\
$^2$Institute for Astronomy, University of Edinburgh, Royal
Observatory, Blackford Hill, Edinburgh, EH9 3HJ, U.K. (SUPA)\\
$^3$University of Nottingham, School of Physics and Astronomy, Nottingham NG7 2RD, U.K.\\
$^4$UK Astronomy Technology Ctr., Royal Observatory, Blackford Hill,
Edinburgh, EH9 3HJ, U.K.\\
$^5$ETH Z\"urich, Institut f\"ur Astronomie, HIT J 11.3,
Wolfgang-Pauli-Str. 27, 8093 Z\"urich
}

\begin{document}
\maketitle
\begin{abstract}

  We present a new method to classify the broad band optical-NIR
  spectral energy distributions (SEDs) of galaxies using three shape
  parameters (super-colours) based on a Principal Component Analysis
  of model SEDs. As well as providing a compact representation of the
  wide variety of SED shapes, the method allows for easy visualisation
  of information loss and biases caused by the incomplete sampling of
  the rest-frame SED as a function of redshift. We apply the method to
  galaxies in the UKIDSS Ultra Deep Survey with $0.9<z<1.2$, and
  confirm our classifications by stacking rest-frame optical spectra
  for a fraction of objects in each class. As well as cleanly
  separating a tight red-sequence from star-forming galaxies, three
  unusual populations are identifiable by their unique colours: very
  dusty star-forming galaxies with high metallicity and old mean
  stellar age; post-starburst galaxies which have formed $\gtrsim$10\%
  of their mass in a recent unsustained starburst event; and
  metal-poor quiescent dwarf galaxies. We find that quiescent galaxies
  account for 45\% of galaxies with $\log{\rm M^*/M_\odot}>11$,
  declining steadily to 13\% at $\log{\rm M^*/M_\odot}=10$.  The
  properties and mass-function of the post-starburst galaxies are
  consistent with a scenario in which gas-rich mergers contribute to
  the growth of the low and intermediate mass range of the red
  sequence.
  
\end{abstract}


%%%%%%%%%%%%%%%%%%%%%%%%%%%%%%%%%%%%%%%%%%%%%%%%%%%%%%%%%%%%%%%%%%
\section{Introduction}
%%%%%%%%%%%%%%%%%%%%%%%%%%%%%%%%%%%%%%%%%%%%%%%%%%%%%%%%%%%%%%%%%%

The extraction of physical properties from the observed spectral
energy distributions (SEDs) of galaxies lies at the heart of gaining a
complete understanding of how galaxies form and evolve.  At low
redshift ($z \lesssim 0.1$), vast numbers of high quality, moderate
resolution spectra have revolutionised our measurement of the physical
properties of galaxies \citep{2000AJ....120.1579Y}. At high redshift
such high quality and extensive spectroscopy remains prohibitively
expensive, but photometric surveys continue to expand at a rapid pace,
in wavelength coverage, area on the sky and depth.

As both observations and spectral synthesis models have improved,
directly observable quantities such as number counts, luminosity
functions and observed-frame colours are commonly replaced with
derived (physical) quantities: mass functions, star formation rates
and rest-frame colours. This process has greatly improved our ability
to compare with galaxy evolution models, and helped us to narrow down
the dominant physical processes leading to the observed galaxy
population over a large fraction of Cosmic time. The
fitting of models to the multiwavelength SEDs of galaxies to obtain physical properties
is now a standard tool in extragalactic astrophysics, with many teams
developing a wide variety of tools, many of which are publicly
available \citep[e.g.][for a recent review]{2008MNRAS.388.1595D,
  Noll:2009p9084, Acquaviva:2011p9004, Walcher:2011p8408}. 

While SED fitting has become the method of choice for estimating the
physical properties of galaxies, it has two primary disadvantages: (1)
it relies on the accuracy of spectral synthesis models
\citep{Conroy:2013p8845}, and (2) galaxies are fit independently from
one another, so excluding additional information provided by the
existence of a population of similar objects.  Colour-colour diagrams
provide a method of displaying large sets of SEDs on two dimensional
diagrams, and classifying galaxies into groups based on observational
(rather than physical) properties. With careful selection of colours
this method can retain most of the physical information in the full
SED \citep[e.g. UVJ][]{Wuyts:2007p8937, Williams:2009p8926}. However,
rest-frame colour-colour diagrams still rely on spectral synthesis
models to perform the K-corrections, while observed-frame
colour-colour diagrams suffer significant degeneracies between
redshift and physical properties for the vast majority of the galaxy
population \citep[e.g.][]{Lane:2007p8846}.

The aim of this paper is to investigate a third method: a rest-frame
(redshift independent) colour-colour diagram that combines the
available information from all the observed bands, but is not
constrained by fits to spectral synthesis models.  We use a Principal
Component Analysis (PCA) of a nominal set of spectral
  synthesis models to calculate the optimal linear combination
of filters needed to separate populations with distinct optical to near
infrared (NIR) SEDs. The
advantages of this method over traditional colour investigations is
that it (i) can determine equivalent linear combinations for galaxies
over a wide redshift range, thus removing the problem of shifting
bands; (ii) reduces the dimensionality of the dataset, eliminating the
need for large numbers of colour-colour diagrams, while providing
colour measurements of higher signal-to-noise ratio (SNR) than can be
obtained from only two magnitudes; (iii) provides observed quantities
which are not constrained to fit the stellar population synthesis
models; (iv) allows easy visualisation of the error incurred on
measuring the true SED shape of galaxies (and thus physical quantities
such as mass-to-light ratio) due to poor sampling of the SED, as a
function of both redshift and SED shape; (v) allows a clear visual
comparison between the SED shapes of observed galaxies and the models
that could be used to subsequently fit for physical parameters, thereby allowing
the identification of populations for which a poor model fit will lead
to incorrect parameter estimation, helping us to understand these
objects rather than removing them from samples due to their bad
$\chi^2$.

The goals of the method presented in this paper are similar in spirit
to the PCA spectral indices developed by \citet{wild_psb}, where the
4000\AA\ region of optical spectra was studied. The PCA identified
three well-known features as being optimal for describing the full
shape of the spectrum (the strength of the 4000\AA\ break, Balmer
absorption lines and the Ca\,H\&K lines). The PCA derived spectral
indices benefit from a greatly enhanced SNR over traditional indices,
by allowing information from all pixels to contribute, including both
multiple correlated lines and the shape of the continuum. The accuracy
and improved visualisation of the complete dataset afforded by these
indices have allowed the identification of post-starburst galaxies
from low resolution Vimos VLT Deep Survey (VVDS) spectra
\citep{Wild:2009p2609} and a complete evolutionary sequence of
starbursts into the post-starburst phase sequence in higher resolution Sloan
Digital Sky Survey (SDSS) spectra \citep{Wild:2010p6226}.
  
The dataset used to test the method is the Ultra Deep Survey (UDS), a
deep, large area near-infrared (NIR) survey and the deepest component
of the UKIRT Infrared Deep Sky Survey \citep[UKIDSS,
][]{2007MNRAS.379.1599L}. The field has deep optical observations from
the Subaru XMM-Newton Deep Survey \citep[SXDS,
][]{Furusawa:2008p8792}, and mid-IR coverage from the Spitzer UDS
Legacy Program (SpUDS, PI:Dunlop). A significant advantage of the
SXDS/UDS field for this study is the large number of homogeneously
observed spectra, taken as part of the UDSz project using a
combination of the VIMOS and FORS2 instruments on the ESO VLT (ESO
Large Programme 180.A-0776, PI: Almaini). These spectra are generally
of insufficient SNR to make robust measurements of
the stellar continuum features for individual galaxies, but can be
stacked to provide a useful verification of the PCA derived
classifications.

This paper is arranged as follows. In Section \ref{sec:pca} we present
the method used to derive the optimal combination of filters to
parameterise the shapes of rest-frame optical-NIR SEDs. We apply the
method to galaxies in the UKIDSS UDS survey at $0.9<z<1.2$ in Section
\ref{sec:uds}, to reveal the physically distinct populations that can
be identified from broad band photometric datasets. We compare with
traditional K-corrected colour-colour and colour-magnitude diagrams,
and stacked rest-frame optical spectra. In Section \ref{sec:results}
we present the stacked SEDs, and the luminosity and mass functions of
the galaxies by class. We use spectral synthesis models to
measure the average physical properties of each class, highlighting
which regions of colour space are not well fit by the models.  Finally,
in Section \ref{sec:disc} we discuss how the method can be used to
compare and contrast different stellar population synthesis models,
investigate the impact of nebular emission lines, and constrain the
dust attenuation law of galaxies. We also calculate the fraction of
true quiescent galaxies as a function of stellar mass at $z\sim1$, and
discuss the implications of our results for using post-starburst
galaxies and low-metallicity quiescent galaxies to improve our
understanding of the evolving galaxy population.

The optimal linear combination of filters (eigenvectors) derived in
this paper are provided in digital format, suitable for surveys using
the same filterset as used here, together with IDL code for reading
and applying them. Where necessary
we assume a cosmology with $\Omega_M=0.3$, $\Omega_\Lambda=0.7$ and
$h=0.7$. All magnitudes are on the AB system. Stellar masses are
calculated assuming a Chabrier IMF and are defined as the stellar mass
at the time of observation: i.e. $M^*=\int {\rm SFR(t)(1-R(t))}
dt$, where R is the fraction of mass in stars returned to the
interstellar medium (ISM) due to supernovae and mass loss.

%%%%%%%%%%%%%%%%%%%%%%%%%%%%%%%%%%%%%%%%%%%%%%%%%%%%%%%%%%%%%%%%%%
\section{Principal Component Analysis of extragalactic photometric
  datasets}\label{sec:pca} 
%%%%%%%%%%%%%%%%%%%%%%%%%%%%%%%%%%%%%%%%%%%%%%%%%%%%%%%%%%%%%%%%%%

Principal Component Analysis and similar techniques have become a
popular tool for QSO and galaxy spectral analysis
\citep{1995AJ....110.1071C, 1998ApJ...492...98G, 2000MNRAS.317..965H,
  2003MNRAS.343..871M, 2004AJ....128.2603Y, 2005MNRAS.358.1083W,
  2006AJ....131...84V, Lu:2006p7218,2007MNRAS.381.1252T, wild_psb,
  Allen:2013p8707}. They serve several purposes, for example, the
exploration of a dataset to uncover unknown trends, the reduction of
dimensionality to speed up model fitting, the optimal separation of
physically distinct components, and the combination of correlated lines
and features into high signal-to-noise ratio spectral indices.
  
These techniques also have great potential for the analysis and
interpretation of photometric datasets, however, their application is
not trivial due to the different redshifts of the galaxies.  Moderate
resolution extragalactic spectra can be de-redshifted to a common
wavelength grid by performing minimal resampling, provided the
redshift range covered by the sample is small enough and the
wavelength range of the spectra is large enough to give sufficient
overlap in wavelength space.  On the other hand, the regions of a
galaxy's rest-frame SED sampled by a photometric dataset depend on the
galaxy's redshift, each filter sums a large number of spectral
features into a single data point, and large wavelength regions can
remain entirely unsampled. For this reason, application of
multivariate statistical techniques to investigate extra-galactic
photometric datasets are limited. Berta et al. (2013) used a Gaussian
Mixture Model to group galaxies with similar SEDs and identify
outliers (following the method presented in Connolly et
al. 2000)\nocite{Berta:2013p8847, Connolly:2000p8848} and
\citet{Dale:2007p7234} performed a PCA on K-corrected colours to
investigate the main variations in galaxy SEDs.

Our aim is to derive an optimal combination of colours (eigenvectors)
that describe the full shape of galaxy SEDs using a minimal number of
variables (principal component amplitudes). Often it is possible to calculate these
optimal combinations from the dataset itself, however, in the case of
extragalactic photometric data this is a complex problem, as
individual galaxies each contribute only a small number of observed
points. We therefore begin our analysis with a large sample of model
SEDs that reasonably cover the colour space of observed galaxies.   While it is important that the models make a reasonable attempt at
  covering the full range of galaxy SED shapes, it is not important
  for them to have the correct relative number density or realistic
  physical properties. From these models we build a large grid of
observed frame colours using all the photometric bands in our survey
and at all redshifts of interest. This will enable us to calculate the
projection of every real galaxy SED onto the eigenvectors regardless of
the redshift of the galaxy or filter coverage.

The models are only used to define the linear combination of filters
that optimally describe the shape of the SEDs. Observed SEDs, or other
models, can be projected onto these linear combinations without any
requirement for the original models to be ``correct''.  Naturally, the
model derived eigenvectors will not be perfectly optimal for the
observed SEDs, in the sense that they will not contain as high a
fraction of the variance of the dataset as for the models from which
they were derived. However, at least in the optical-to-NIR wavelength
range studied here, galaxy SED shapes are well enough understood that
eigenvectors derived from any reasonable set of models or real data
will be very similar. The approach described here forces a complete separation
between observables (galaxy SED shapes) and model derived properties
(mass-to-light ratios, dust contents, metallicities etc.).

\subsection{Building a super-sampled colour grid}

\begin{figure}
  \includegraphics[scale=1]{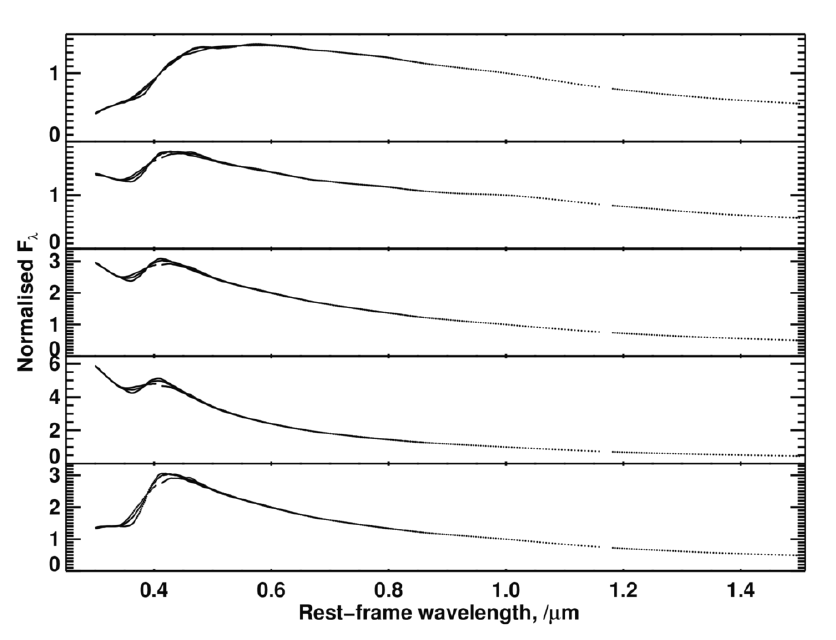}
  \caption{Examples of super-sampled model SEDs used to derive the
    optimal linear combination of filters needed to describe the shape
    of galaxy SEDs. Each point represents the normalised flux of the
    model galaxy observed with one filter at one redshift, placed at
    the effective rest-frame wavelength of the filter. All SEDs are
    normalised at 1\mum. Examples have been chosen to illustrate the
    full range of model galaxies: in the top four panels
    light-weighted mean stellar age decreases from top to bottom. The
    bottom panel shows a post-starburst model galaxy.
  }  \label{fig:mockspec}
\end{figure}

We begin with a library of $\sim$44,000 ``stochastic burst'' model
SEDs built from \citet{2003MNRAS.344.1000B} stellar population
synthesis models with stochastic star formation histories:
exponentially declining star-formation rates with inverse decay rates
between 0 and 1\,Gyr$^{-1}$, with $\delta$-function bursts of star
formation randomly superposed with varying strengths and ages. The
models have a flat prior on metallicities between one tenth and twice
solar, and a flat prior on formation time between 0 and 6\,Gyr (the
age of the Universe at the lowest redshift we are interested in). Dust
attenuation is included using the \citet{2000ApJ...539..718C}
prescription, where stars older than $10^7$ years suffer a fixed
fraction ($\mu\sim0.3$) of the total effective attenuation
($\tau_V$). The models have a Gaussian prior on the total effective
dust attenuation centred on $\tau_V=1$, which leads to between 0 and 6
magnitudes of attenuation in the $V$-band for stars younger than
$10^7$ years, and between 0 and 1.8 magnitudes of attenuation for
older stars. Similar libraries have been widely used in the literature
to determine the physical properties of galaxies
\citep{2003MNRAS.341...33K, Gallazzi:2005p6450, Salim:2007p6062,
  2008MNRAS.388.1595D}.  Altering the priors for the model grid
  was found to have no significant effect on the output from the PCA,
  so long as a reasonable range of SED shapes was included. Ultimately
  this occurs because a limited number of physical processes
  contribute to the optical-to-NIR SED of galaxies. In this wavelength
  range the origin of the SED shapes is reasonably well understood,
  we therefore do not expect the use of different spectral synthesis
  models to have a significant impact on the first few eigenvectors.
  Obtaining such a robust result for ultraviolet or far-infrared wavelength regions
  would be more challenging. 

Nebula emission lines are not included in the models, as this remains
an uncertain procedure. We tested the effect of their inclusion using
a limited set of the strongest lines, with amplitude based on the
number of ionising photons output from the stellar population
synthesis model, and standard line ratios for star forming galaxies.
Their addition was found not to have a significant effect on the
eigenvectors. Emission line contamination in the data may subsequently
be identified as regions of colour space where the models and data
disagree (see Section \ref{sec:emlines}). We emphasise that these
stochastic burst models do not need to be a perfect representation of
the data: any models can be projected onto the eigenvectors if they are 
subsequently deemed more appropriate to determine physical parameters
from the data. 

We convolve these model SEDs with the Subaru XMM-Newton Deep Survey
(SXDS), UKIDSS ultra deep survey (UDS) and Spitzer UDS Legacy Program
(SpUDS) filter sets (Subaru $B$, $V$, $R$, $i'$, $z'$; UKIRT$J$, $H$,
$K$; SpUDS IRAC 3.6\mum\ and 4.5\mum), progressively shifting the
model SEDs to redshifts of $0.9<z<2.0$ in steps of $ \Delta z =
0.01$. Additionally the ESO $Y$-band, HST 125W (J) and 160W (H)
filters are included as placeholders, but not used in the analysis
presented in this paper.  Smaller redshift steps were tested but led
to no improvement. This redshift range was chosen to be physically
interesting (where the majority of stellar mass in the Universe is
formed), and also to include redshift regimes where the SED shapes are
both well and less-well constrained. Any redshift range could be
chosen depending on the science question being asked. The resulting
flux matrix is truncated at rest wavelengths of between 3000\AA\ and
1.5\mum, to avoid more uncertain regions in spectral synthesis
models, giving a ``super-sampled'' matrix of 935 data points. Some
examples of model galaxy SEDs filtered in this way are shown in Figure
\ref{fig:mockspec}. Each filter at each redshift is represented by a
dot placed at the effective \emph{rest-frame} wavelength of the
filter. Gaps indicate rest-frame wavelength regions that are not
covered by any filter at any redshift. Conversely multiple traces
appear in some regions, where two or more different filters sample the
same wavelength region for galaxies at different redshifts.

\begin{figure}
  \includegraphics[scale=1]{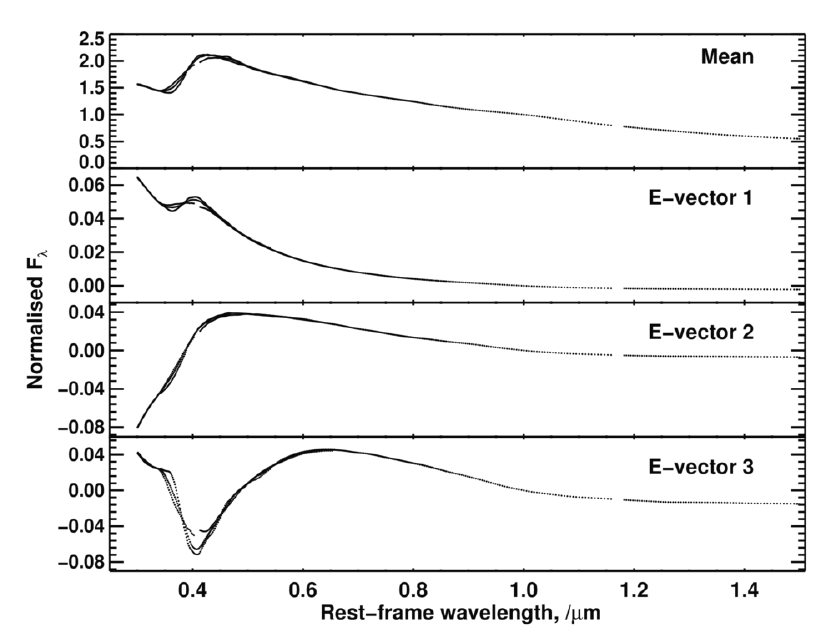}
  \caption{The mean and first 3 eigenvectors from a Principal
    Component Analysis of a library of Bruzual \& Charlot (2003)
    ``stochastic burst'' model SEDs. Each point represents the
    normalised flux observed with one filter at one redshift, placed
    at the effective rest-frame wavelength of the filter. On addition
    to the mean, the first eigenvector primarily alters the red-blue
    slope of the SED, the second and third primarily alter the shape
    of the 4000\AA/Balmer break region.} \label{fig:evecs}
\end{figure}

We normalise all model SEDs at 1\mum, calculate the mean flux array,
subtract this from each model SED, and perform a PCA of the
resulting difference array i.e. we calculate the eigenvectors of the
covariance matrix of the difference array.  Both a narrative and
mathematical description of our application of PCA can be found in
Section 3 and Appendix A of \citet{wild_psb}\footnote{See
  \citet{1987mda..book.....M} for an introduction to the mathematics
  of PCA. IDL code for performing PCA and attempts to homogenise
  different formalisms, is available for download from
  http://www-star.st-and.ac.uk/$\sim$vw8/downloads.  Two similar
  multivariate analyses were attempted, non-negative matrix
  factorisation and independent component analysis (ICA), however, PCA
  provided the most satisfactory results when applied to noisy and
  gappy data. }.  The first 3 eigenvectors are shown in Figure \ref
{fig:evecs}, and depict how the flux at different wavelengths is
correlated within the set of model galaxy SEDs.  By construction, the
eigenvectors are ordered by the variance that they account for in the
input dataset: the first 3 eigenvectors represent 96.9\%, 2.9\% and
0.15\% of the variance, 99.98\% in total.

A normalised galaxy SED ($f_\lambda/n$) can be approximately
reconstructed from the mean spectrum ($m_\lambda$) plus a linear
combination of the first $p$ eigenvectors ($e_{i\lambda}$):
\begin{equation}\label{eqn:pca1}
\frac{f_\lambda}{n} = m_\lambda + \sum_{i=1}^p a_i e_{i\lambda}.
\end{equation}
In perfect data, the accuracy of the reconstruction increases with
increasing numbers of eigenvectors used.  The linear combination
coefficients ($a_i$), also known as principal component amplitudes,
uniquely define the shape of a galaxy's SED and can be calculated for
any arbitrary galaxy SED by inverting Eqn. \ref{eqn:pca1}:
\begin{equation}\label{eqn:pca2}
a_i = \sum_\lambda \frac{f_\lambda}{n} e_{i\lambda}. 
\end{equation}
The first three principal component amplitudes alone provide a compact
representation of the shape of all the input SEDs, accounting for
99.98\% of the variance in SED shapes in our model dataset.
``Colours'' are the simplest linear combination of filters that can be
used to describe the shape of galaxy SEDs. From now on we will refer
to the principal component amplitudes as ``super-colours'' (SCs): while colours
are the combination of two filters weighted equally, super-colours
combine multiple filters with an optimally defined weighting scheme
(the eigenvectors).

The super-colours can be thought of as the ``amount'' of each
eigenvector contained within each galaxy SED.  The mean SED is an
averagely blue galaxy (top panel of Figure \ref{fig:evecs}), which can
be made bluer or redder by adding or subtracting some of the first
eigenvector (second panel).  The second eigenvector alters the
strength of the 4000\AA\ or Balmer break.  The third eigenvector is
also associated with the shape of the SED around 4000\AA. This tells
us that to first order a model galaxy SED can be described by its
overall slope (colour); to second order the exact shape of the
4000\AA\ break region is important. The PCA formally quantifies these
well known properties of galaxy SEDs.

\begin{figure*}
\includegraphics[scale=1]{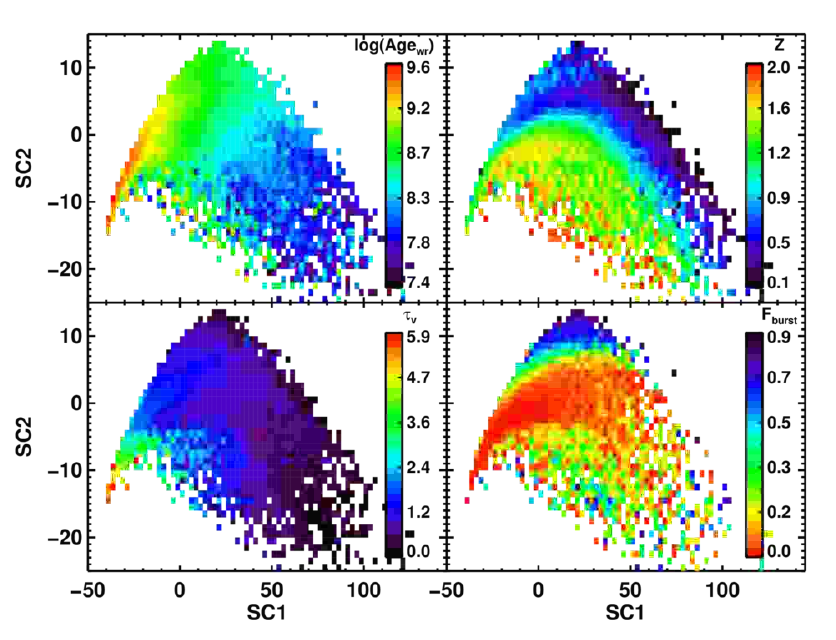}
\includegraphics[scale=1]{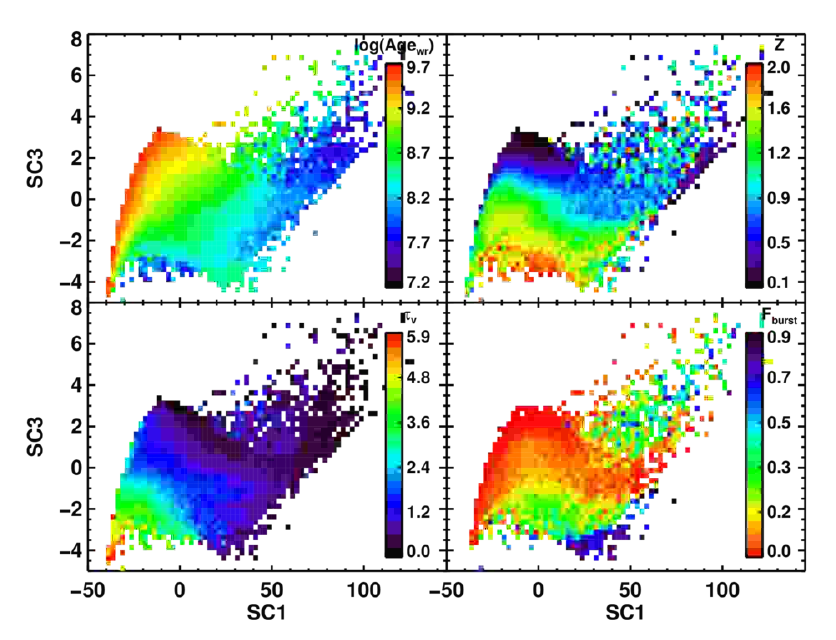}
\caption{Super-colour (SC) diagrams coded by the mean physical
  properties of the model galaxies in each bin, for SC1/2 (left) and
  SC1/3 (right). From top to bottom and left to right: log of r-band
  light weighted mean stellar age (yr), metallicity (relative to
  solar), total effective $V$-band optical depth affecting stars
  younger than $10^7$ years (=0.92$A_V$, where $A$ is attenuation in
  magnitudes), fraction of stellar mass formed in bursts in the last
  Gyr.  }\label{fig:params}
\end{figure*}

In Figure \ref{fig:params} the super-colours of the $\sim$44,000
stochastic burst model SEDs are shown, with regions of super-colour
space colour coded by the mean physical parameters of the models lying
in that region. From top left to bottom right we show the $r$-band
light-weighted mean stellar age, metallicity, total effective dust
attenuation and fraction of stars formed in starbursts in the last
1\,Gyr. As expected, the first super-colour (SC1) correlates primarily
with both mean stellar age and dust content -- this is a well known
degeneracy in photometric studies of galaxies. Both the second and
third super-colours (SC2 and SC3) correlate with metallicity, and provide
additional information on age and dust, which can break the degeneracy
between these parameters in some cases. For example, very highly
attenuated, high metallicity, old galaxies are identifiable at low SC1,
SC2 and SC3; and old, low-metallicity galaxies can be isolated at
low SC1 and high SC3. Another interesting region of colour space is the population
of mixed metallicity objects at high SC2, which contains galaxies
that have undergone a significant burst of star formation in the last
1\,Gyr (post-starburst galaxies). 

%------------------------------------------------------------------
\subsection{The reality of sparse sampling}\label{sec:sparse}
%------------------------------------------------------------------
\begin{figure}
\includegraphics[scale=1]{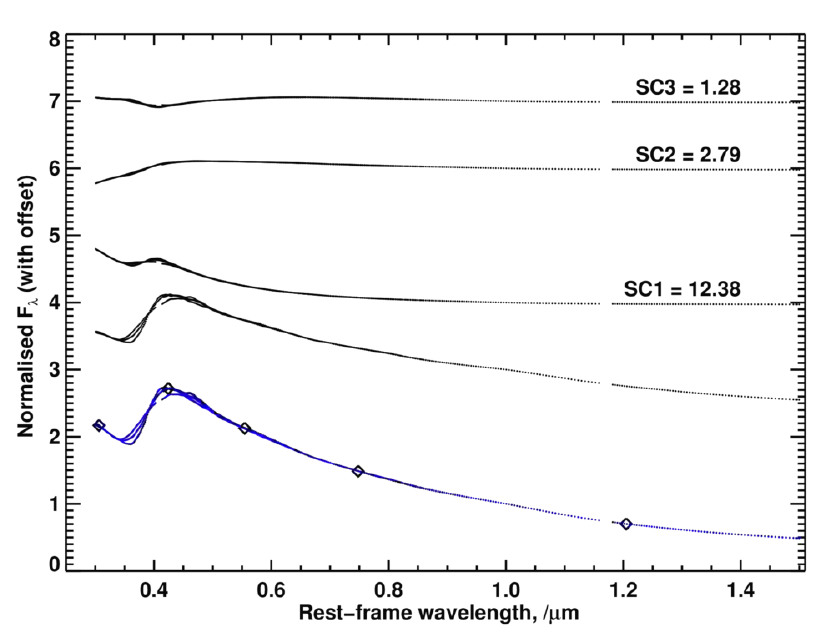}
\caption{An example of how a model SED ``observed'' at a redshift of
  1.915 with 5 photometric bands (diamonds) can be reconstructed from
  the mean and first three eigenvectors. The amplitude of the
  super-colours measured for this galaxy are given above the
  eigenvectors. The reconstructed SED (blue) is almost
  indistinguishable from the underlying input SED (black). A vertical
  offset has been applied to the mean and eigenvectors for
  clarity.}\label{fig:egpca}
\end{figure}

\begin{figure}
\includegraphics[scale=1]{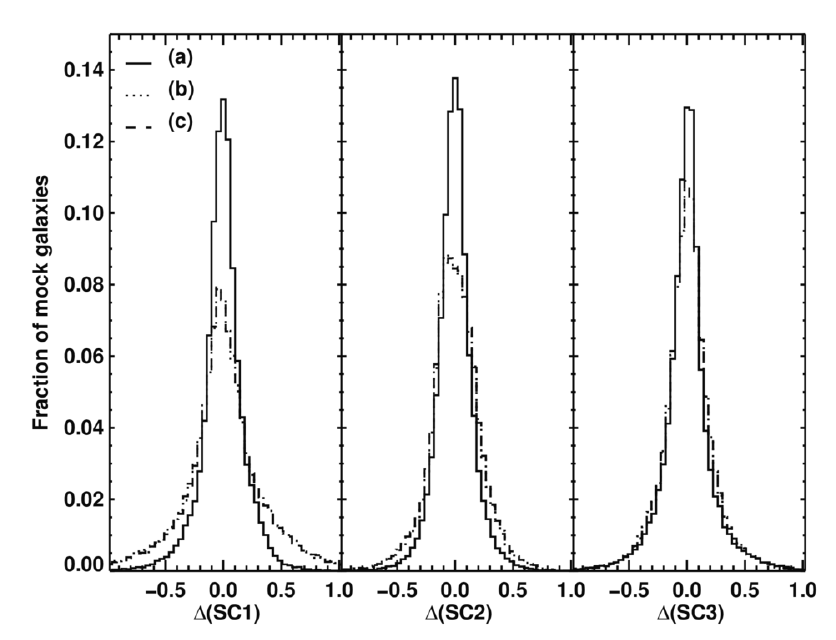}
\caption{The distribution of errors on the super-colours caused by
  incomplete and inaccurate information (see text for more details on
  each scenario): (a) the SED is sparse-sampled but the normalisation
  is known; (b) the SED is sparse-sampled with unknown flux normalisation;
  (c) additionally including redshift errors. During these tests,
  model galaxies are assigned random redshifts between 0.9 and 2.0 and
  ``observed'' with the SXDS/UDS filter set.  Redshift errors were
  drawn from a Gaussian distribution with $\sigma_z=0.06$. }
\label{fig:deltapcs}
\end{figure}

Up to this stage, we have shown how the principal component amplitudes of a
photometric dataset can be thought of as ``super-colours'' which
provide a compact representation of the shape of galaxy SEDs. We have
not yet addressed the core problem: in reality, each galaxy is
observed at a single redshift, with only a handful of filters covering
the wavelength range of interest. To demonstrate the effect of
``sparse sampling'' the array, we assign each of our $\sim$44,000
stochastic burst model SEDs a random redshift in the range used to
build the super-sampled array ($0.9<z<2.0$). Each SED is observed with
between 5 and 7 filters, depending on redshift, that partially cover
the full wavelength range. We project this sparsely sampled data
vector onto the eigenvectors, using the ``gappy-PCA'' technique of
\citet{1999AJ....117.2052C}, weighting the observed bins by
$1/\sigma^2$ and the unobserved bins by zero, where $\sigma$ is the
error on the observed data point.  This technique estimates the
principal component amplitudes in the presence of gappy and noisy data by
minimising the error in the reconstructed spectrum, over the full
range in wavelength, weighted by the variance of the observed bins
(filters). In this mock dataset we set $\sigma$ to be constant in all
observed bins. An additional effect caused by the sparse-sampling of
the data is that the precise flux normalisation of the SED is unknown
and must be fitted for simultaneously with the super-colour amplitudes
\footnote{The normalised-gappy PCA algorithm was created by G. Lemson (MPA)
  and IDL code is available for download from
  http://www-star.st-and.ac.uk/$\sim$vw8/download} \citep{wild_psb}.
Figure \ref{fig:egpca} shows an example of how a sparse-sampled SED is
reconstructed from the eigenvectors and the three measured
super-colour amplitudes.

\begin{figure*}
\includegraphics[scale=1]{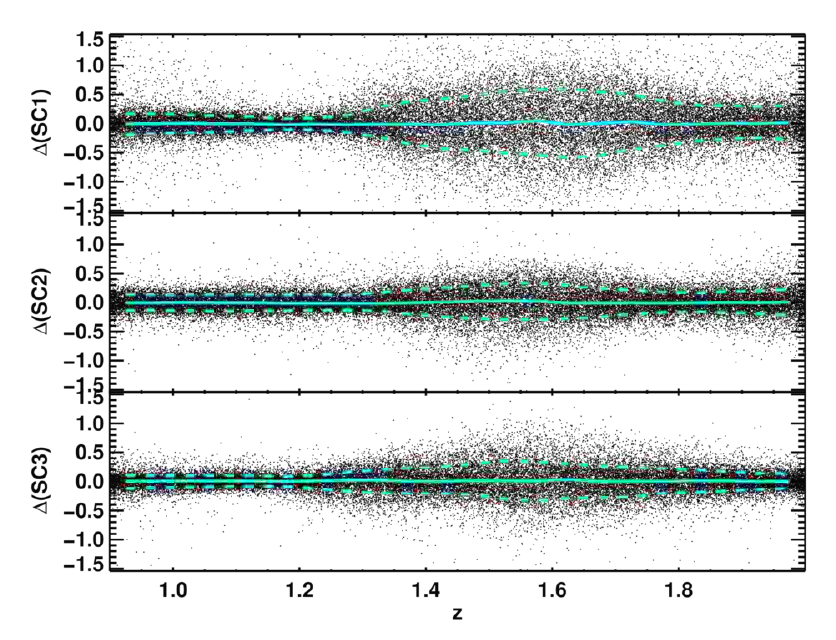}
\includegraphics[scale=1]{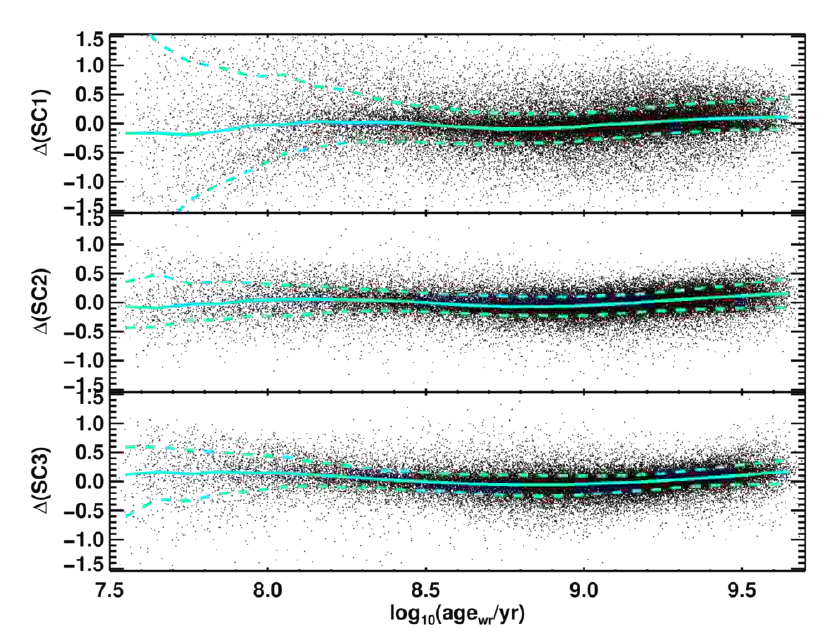}
\caption{Errors on the super-colours caused by sparse-sampling,
  unknown normalisation and redshift errors, as a function of redshift
  and mean stellar age of the SED. During these tests, model galaxies
  are assigned random redshifts between 0.9 and 2.0 and ``observed''
  with the SXDS/UDS filter set.  Redshift errors were drawn from a Gaussian
  distribution with $\sigma_z=0.06$. Overplotted lines indicate the 
  16th, 50th and 84th percentiles of the distributions.
}\label{fig:deltapcs_trend}
\end{figure*}

The accuracy to which the input SED can be recovered will depend on
the filter set, the galaxy redshift and shape of the SED.  In Figure
\ref{fig:deltapcs} we show the error on the super-colours for our mock
catalogue of $\sim$44,000 stochastic burst model SEDs, sparse-sampled
with redshifts randomly assigned in the range $0.9<z<2.0$. The solid
line (a) compares the super-colours measured from the super-sampled
and sparse-sampled array, but assuming that we know the flux
normalisation\footnote{This will never be the case with observed
  galaxy SEDs, but is included to emphasise the importance of fitting
  for the unknown normalisation as an additional step in solving for
  the principal component amplitudes.}. The dotted line (b) includes the effect
of unknown flux normalisation, and the dashed line (c) shows the
effect of errors on the redshifts. To model the effect of photometric
redshifts we assume a normal
distribution of redshift errors, with a $1\sigma$ width of 0.06
(percentage error of 7\% and 3\% at $z=0.9$ and $2.0$).\footnote{We
  take a fixed width for the distribution of redshift errors, which is
  the variance of the distribution of redshift error for 282 galaxies
  with spectra presented in Section \ref{sec:spec}. In reality
  photometric redshift errors depend on redshift, and catastrophic
  outliers exist. We do not attempt to model all eventualities here,
  which will depend on the dataset being studied.} It is clear
  that both the sparse-sampled nature of the data and the unknown
  normalisation affect our ability to recover the SED shape, however,
  redshift errors at the level found in photometric datasets have no further significant impact
  on the measured super-colours. The 16th and 84th percentiles of the
distributions (equivalent to 1$\sigma$ errors for a Gaussian
distribution) when all effects are included (histogram c of Figure \ref{fig:deltapcs}) are 0.28 and
0.31 for SC1, 0.18 and 0.19 for SC2, and 0.17 and 0.18 for SC3. There
are no systematic offsets.

In Figure \ref{fig:deltapcs_trend} we show the errors on the
super-colours when all effects are included (histogram c), as a
function of redshift and SED shape (parameterised by the
light-weighted mean stellar age of the model). For the 
available filters in the UDS we obtain accurate super-colours, and can
therefore reliably recover the SED shapes, in the redshift ranges of
$0.9<z<1.2$ and $z>1.7$. In the low-redshift range ($0.9<z<1.2$), the
16th and 84th percentiles are 0.14 and 0.13 for SC1, 0.14 and 0.14 for
SC2, and 0.09 and 0.1 for SC3. The percentage errors are 1.2\% for
SC1, 7\% for SC2 and 15.8\% for SC3.  This figure highlights how,
  for certain redshift ranges, the ability to constrain the shape of a
  galaxy SED is significantly impaired by gaps in wavelength coverage
  of the broad band filters. Clearly, the physical parameters
  determined from SED fits of galaxies at $1.2<z<1.7$ observed with
  the UDS filter set will be less well constrained than at other redshifts. 

The younger SEDs suffer larger errors, due to the lack of strong
features in their SED. The older SEDs suffer from a systematic offset
in SC1 and SC2, at the level of about 1\%. This is unlikely to
be a dominant source of error in the analysis of a real dataset, but
is worth keeping in mind if highly accurate ages of old stellar
populations are required from SED fitting.

\begin{figure}
\includegraphics[scale=1]{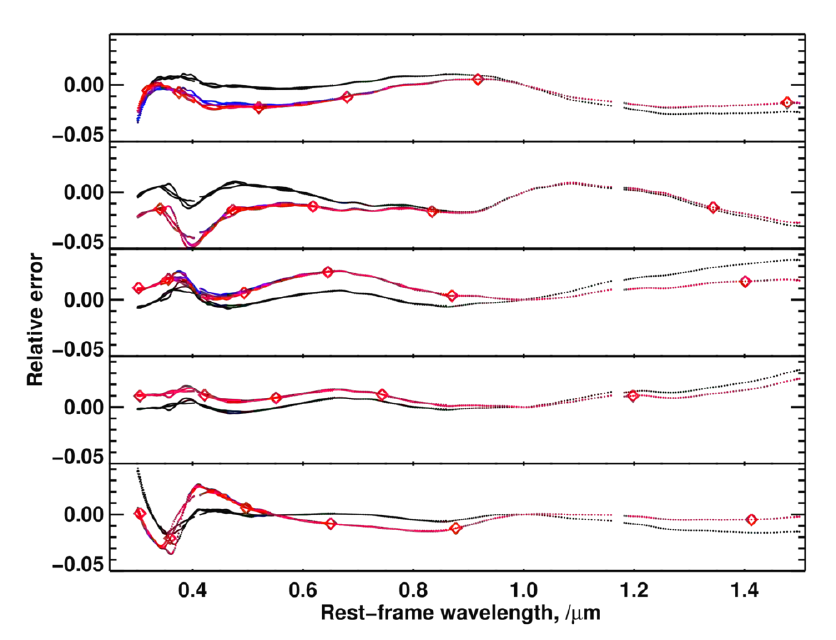}
\caption{The fractional error between the true and recovered SED for
  the examples shown in Fig. \ref{fig:mockspec}. The black trace shows
  the effect of only using three eigenvectors, the blue trace includes
  the effect of sparse-sampling and unknown flux normalisation, and
  the red trace additionally includes the effect of redshift errors.
  The red trace sits on top of the blue trace, showing that
  photometric redshift errors do not significantly impair our ability
  to measure SED shapes. From top to bottom the redshifts of the
  mock galaxies are 1.50, 1.63, 1.47, 1.97, 1.49. The red diamonds
  indicate the effective wavelengths of the filters used in the
  sparse-sampled cases.}\label{fig:mockrecon}
\end{figure}

In Figure \ref{fig:mockrecon} we show examples of the fractional error
between the input and recovered SEDs, for the same five randomly
selected model SEDs as shown in Figure \ref{fig:mockspec}. The black trace
shows the effect of only using three eigenvectors, the blue trace
includes the effect of sparse-sampling and unknown flux normalisation
and the red trace additionally includes the effect of redshift
errors. The blue trace is not easily visible under the red trace, which
again emphasises the fact that photometric redshift errors have little additional
impact on our ability to recover the SED shape.

%%%%%%%%%%%%%%%%%%%%%%%%%%%%%%%%%%%%%%%%%%%%%%%%%%%%%%%%%%%%%%%%%%
\section{Application to galaxies in the SXDS / UDS} \label{sec:uds}
%%%%%%%%%%%%%%%%%%%%%%%%%%%%%%%%%%%%%%%%%%%%%%%%%%%%%%%%%%%%%%%%%%

\begin{table}
   \begin{center}
     \caption{$5\sigma$ limiting depths in AB magnitudes (3\arcsec\ aperture) of the SXDS (optical) and UDS
       (near-IR) photometry used to measure the super-colours of
       $0.9<z<1.2$ galaxies, and the effective
       wavelengths of the filters \citep{1996AJ....111.1748F}. }
\label{tab:depths}
     \begin{tabular}{ccccccc}\hline\hline
Band & $R$ & $i'$ & $z'$ & $J$ & $H$ & $K$ \\
       \hline
Depth & 26.4 & 26.3 & 25.6 & 24.9 & 24.4 & 24.6\\
$\lambda_{\rm eff}$($\mu m$)  & 0.6507 &  0.7646 & 0.9011 & 1.2483 &
1.6319 & 2.2010 \\
       \hline
     \end{tabular}
   \end{center}
 \end{table}

 The Ultra Deep Survey (UDS) is a deep, large area near-infrared (NIR)
 survey centred on RA = 02:17:48, DEC = -05:05:57 with deep optical
 observations from the Subaru XMM-Newton Deep Survey \citep[SXDS,
 ][]{Furusawa:2008p8792}, and mid-IR coverage from the Spitzer UDS
 Legacy Program (SpUDS, PI:Dunlop). The effective wavelengths and
 depths of the observations for the filters used to calculate the
 super-colours of $0.9<z<1.2$ galaxies in this paper are given in Table
 \ref{tab:depths}.  We use a $K$-band selected catalogue based on the
 eighth UDS data release (DR8), but update the photometry to DR10 and
 include new $z'$-band observations (Furusawa et al. in preparation,
 Bowler et al. 2012). The total survey area with full optical, NIR and
 mid-IR coverage and with conservative masking of diffraction spikes
 is 0.607\,deg$^2$.

 Photometric redshifts were calculated by fitting the observed
 photometry ($BVRi'z'JHK$ and 3.6 and 4.5\mum\ when available) with
 synthetic and empirical galaxy templates using code based on the
 public package HYPERZ \citep{Bolzonella:2000p8741}. Further details
 are given in \citet{Cirasuolo:2007p8739} and
 \citet{Cirasuolo:2010p8794}. While it would be possible to solve for
 both the super-colours and photometric redshifts in a single step, we
 decided to focus on recovering the SED shapes in this paper, and use
 the well tested photometric redshifts calculated from a standard code
 with the full set of available filters.

 We select 39,683 sources with $K<23$, of which 6,912 lie in the
 redshift range $0.9<z_{phot}<1.2$.  We select this redshift range for
 this initial analysis for four reasons: (1) the errors on the
 super-colours due to sparse-sampling of the SEDs is minimal given the
 current set of filters available in the UDS field (see Section
 \ref{sec:sparse}); (2) there are a large number of spectra with
 coverage of the 4000\AA\ rest-frame region that are available to test
 our method (see Section \ref{sec:spec}); (3) the strongest emission
 lines do not intercept narrow filters which may cause biases in the
 measured super-colours; (4) we do not require the IRAC photometry
 which is less reliable than the optical and NIR photometry.

 We calculate the super-colours for the observed galaxies in exactly
 the same way as for the model galaxies (Eqn. \ref{eqn:pca2}), with
 errors derived from the depths in each band (Table
 \ref{tab:depths}). Bins with no information are given zero weight. In
 the following subsections we include all galaxies, regardless of the
 formal $\chi^2$ value for the photometric redshift fit or the
 distance between the observed super-colours and closest
 model galaxy. Excluding objects based on a goodness-of-fit to models can
 lead to biases in the samples when the models are not a perfect
 representation of the data.

%------------------------------------------------------------------
\subsection{Super-colours: data vs. models}\label{sec:classes}
%------------------------------------------------------------------

\begin{figure*}
\includegraphics[scale=1]{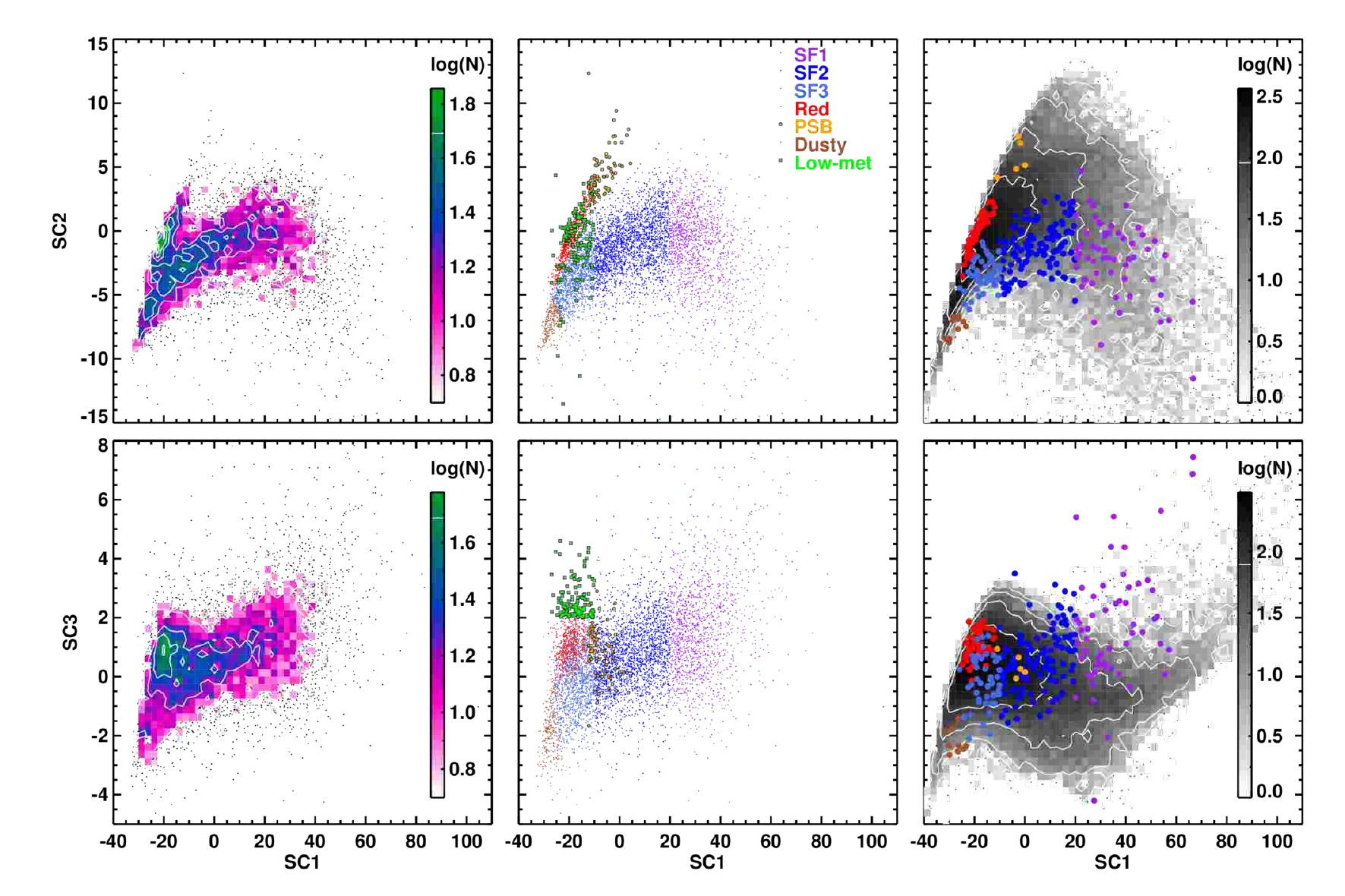}
\caption{Super-colour (SC) diagram for galaxies in the UDS field, for
  SC1 vs. SC2 (top) and SC1 vs. SC3 (bottom).  The tight red-sequence
  can be observed in the upper-left of SC1/2, with the blue-cloud
  extending from bottom left to high values of SC1.  \emph{Left:}
  Colour scale indicates number density, with individual galaxies
  included as dots only in lower density regions of the diagram.
  \emph{Centre:} all galaxies are plotted as dots or symbols, colour
  coded by our nominal classification scheme (see text): red-sequence
  (red), blue-cloud (with decreasing mean stellar age from
  cyan to blue to purple, [with arbitrary boundaries]), post-starburst
  (larger orange circles, boundary determined from comparison to spectra),
  dusty-starforming (brown), low-metallicity (larger green squares). \emph{Right:} the
  greyscale shows the distribution of stochastic burst model galaxy
  colours (black dots in low density regions), with the subsample of
  UDS galaxies with UDSz spectra overplotted and colour coded to allow
  direct comparison.}
\label{fig:pc123}
\end{figure*}

In Figure \ref{fig:pc123} we show the distribution of the first three
super-colours for the UDS galaxies, using both a number density
representation (left panel) and plotting individual points (central
panel). To facilitate comparison between the stochastic burst models
used to build the eigenvectors and the data, we show the distribution
of model colours in the right-hand panel. In the right panel we
overplot the position of UDS galaxies which have spectra (Section
\ref{sec:spec}), colour coded
according to their position in super-colour space.

Although there are no truly discrete populations, separation of
colour-colour space into classes will allow us to to stack galaxies
with similar SEDs to study their average properties in more detail.
Comparing to the mean model parameters shown in Figure
\ref{fig:params} we can identify a tight red-sequence to the
upper-left in SC1 vs. SC2.  The more diffuse blue-cloud spreads across
the full range of SC1, with mean stellar age decreasing with
increasing SC1. We colour code the red-sequence galaxies as red and
split the blue-cloud into three classes from cyan to blue to purple
with decreasing mean stellar age. The position of these blue-cloud
boundaries is arbitrary.  

More unusual classes of galaxies can also be isolated. Galaxies that
lie below and to the left of the red-sequence in SC1/2 and SC1/3 are
expected to have very high dust contents; we colour code them brown to
distinguish them from normal blue-cloud or red-sequence galaxies.  Galaxies
that have had a recent burst of star formation (post-starburst) form a
distinct sequence entering the top-right of the red-sequence in SC1
vs. SC2; they are colour coded orange. The boundary between the red
and post-starburst populations is determined empirically from
comparison to the spectra (see Section \ref{sec:spec} and Appendix \ref{app:vvds}) to
ensure that the photometric selection is as close as possible to a
traditional spectroscopic selection. Finally, we identify a population
of red-sequence galaxies with high SC3 (green points), which are
consistent with low-metallicity red-sequence galaxies. 

The stochastic burst model stellar populations shown in the right hand
panel are not intended to reproduce the relative number density of
different populations, however, their wide range of star formation
histories has been constructed to allow full coverage of the colour
distribution of real galaxies as far as possible. There is no a-priori
reason that the data and models should lie in the same region of
colour-colour space, as the data is not forced to fit the models as is
the case when traditional K-corrections are performed. Differences
between model and data colour distributions could occur for many
reasons: incorrect star formation histories of the models; incomplete
or incorrect stellar population models (lack of particular types of
stars, incorrect evolutionary tracks); incomplete or incorrect model
SEDs (wrong assumed dust attenuation, lack of emission lines);
zero-point offsets in the photometry; photometric redshift errors.
The fact that the observed galaxies are largely contained within the
model distribution indicates that the  spectral synthesis models
  used to build the eigenvectors are sufficiently accurate to reproduce the spectral
energy distributions of most galaxies at $z\sim1$.

Looked at another way, this figure shows that current rest-frame
optical-NIR photometric observations of the vast majority of $z\sim1$
galaxies can not drive further substantial improvement in the models.
However, there are three small offsets between model and data which suggest
where improvements in the models might be made: the dustiest class has
slightly higher SC1 values than is possible with the models; the
youngest blue-sequence galaxies have higher SC3 values than is
typically seen in the models; there is a population of red objects
with very high SC3 values. In Section \ref{sec:disc} we will
investigate the reasons for each of these offsets with respect to
possible observational errors or improvements in the population
synthesis models.

There are a small number of galaxies with super-colours that scatter
away from the main sample. All objects have been included in this
plot, regardless of the goodness-of-fit of the first three
eigenvectors to the observed points, or the formal errors on the
principal component amplitudes or photometric redshifts. Two interesting
populations appeared during our investigations of the
outliers. Firstly, some of the outliers falling below the
blue-sequence have X-ray detections, or are identified as
spectroscopic broad line AGN (Ueda et al. 2008; Simpson et al. 2012; Akiyama et al. in prep.)\nocite{2008ApJS..179..124U,2012MNRAS.421.3060S}. The strong UV continuum from
the AGN causes the objects to have blue SEDs and acts to reduce the
strength of the Balmer break (or obliterate it entirely in many
cases). Secondly, the output from some photometric redshift catalogues
tested during initial studies for this paper caused a large scatter of
objects to the upper right (small SC1 and large SC2). This was found
to be caused by ``aliasing'' effects, whereby galaxies close to a
certain redshift are all forced onto a similar, but incorrect
redshift, due to the presence of a strong discontinuity in the
SED.  While we have verified that the photometric redshift
  catalogue used in this paper does not suffer from strong aliasing
  biases, photometric redshift errors remain a leading candidate for
  causing the outliers in super-colour space.

%------------------------------------------------------------------
\subsection{Comparison to medium resolution spectra}\label{sec:spec}
%------------------------------------------------------------------

\begin{figure*}
\includegraphics[scale=0.85]{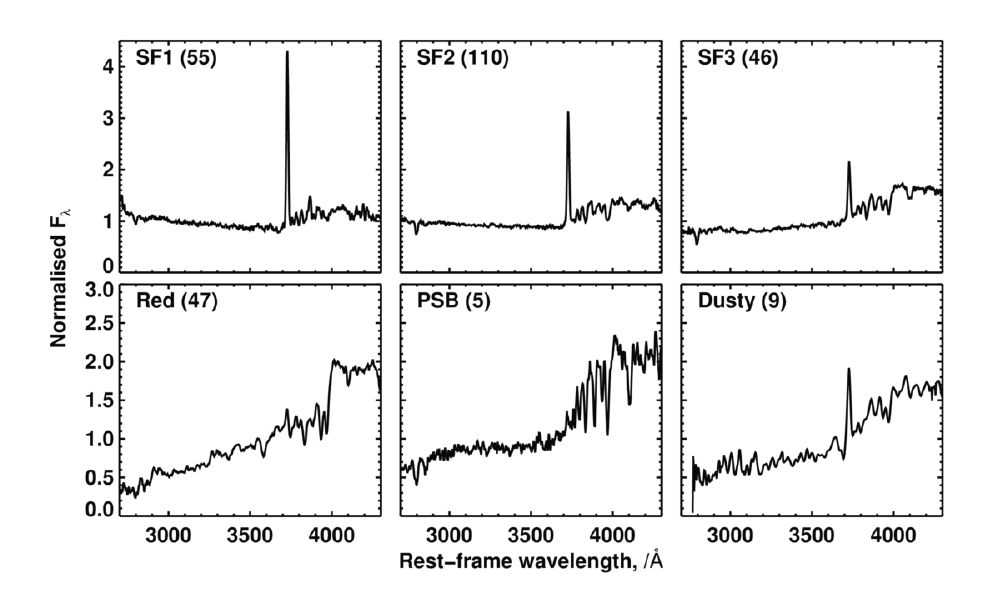}
\includegraphics[scale=0.85]{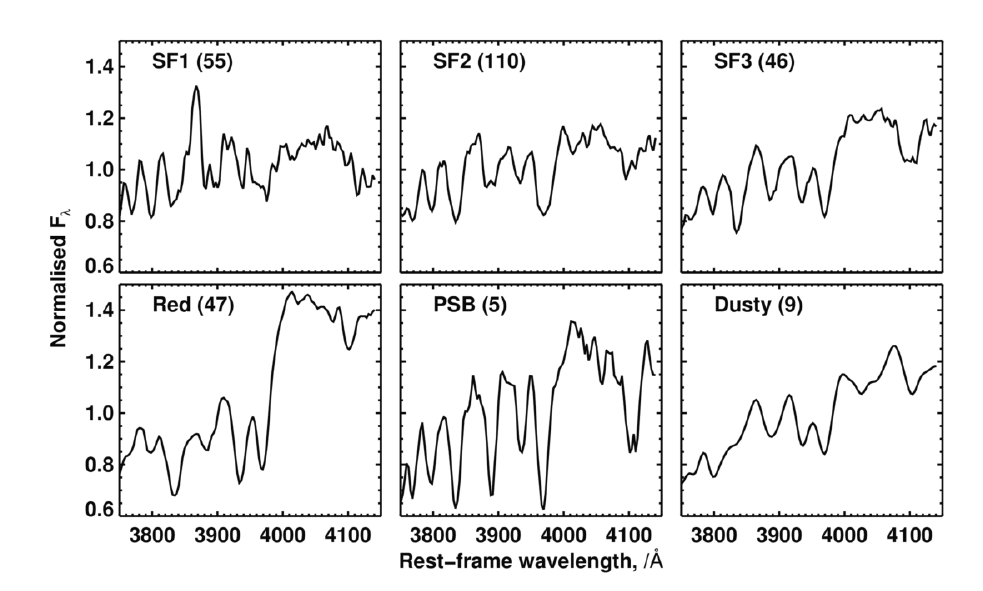}
\caption{Stacked UDSz spectra from each of the super-colour defined
  galaxy classes as indicated in the top left of each panel. The
  number of galaxies included in each stack is also given. On
  the left the full wavelength range is plotted, on the right we zoom in
  on the 4000\AA\ region.  }\label{fig:stacks}
\end{figure*}

A significant advantage of the SXDS/UDS field for this first
super-colour study is the large number of homogeneously observed
spectra, taken as part of the UDSz project (ESO Large Programme
180.A-0776, PI: Almaini) using a combination of the VIMOS and FORS2
instruments on the ESO VLT. 2,881 optically brighter systems ($i'<24$
or $V<25$) were targeted with the VIMOS LR Blue grism (4.5 hours
exposure) and LR Red grism (2.6 hours exposure), with a spectral
resolution of $R$ = 180 and 210 respectively. 802 fainter, redder
systems ($i'< 25$ and $V > 24$) were targeted with the FORS2 300I
grism (5.5 hour exposure), with a spectral resolution of $R$=660.
Details of the reduction are given in \citet{Bradshaw:2013p8772} for
VIMOS and \citet{CurtisLake:2012p8773} and \citet{McLure:2013p8908}
for FORS2.

From the complete catalogue we select 752 galaxies with secure
redshifts and photometric matches in our masked UDS field, of which
335 lie in the redshift range $0.9<z_{\rm phot}<1.2$. We further
remove objects with spectral per-pixel SNR$<2$ and rest-frame
wavelength coverage that does not extend red enough to include
4150\AA. This results in a final sample of 282 moderate quality
spectra with coverage of the 4000\AA\ break region. The spectra are
not all of high enough quality to provide useful stellar continuum
information for individual objects, but they can be stacked to confirm
our super-colour based SED classifications. For the purposes of
spectral stacking, we convolved the FORS2 spectra to a resolution of
$R=200$ to match the VIMOS spectra and rebinned all spectra onto a
common wavelength scale using a linear interpolation. The stacks were
built using an arithmetic mean, with each spectrum normalised by its
median flux. A weighted mean was also calculated, but did not improve
results.

The availability of rest-frame optical spectra for galaxies in most of
our super-colour defined classes provides an easy way to verify the
method. One advantage of optical spectra over broad band photometry is
that dust does not substantially change the strength of the 4000\AA\
break or absorption lines, and therefore the age of the stellar
population can be ascertained much more accurately.  In Figure
\ref{fig:stacks} we show stacked spectra of galaxies in each
super-colour class, over the full wavelength range (left panel) and
zooming in on the 4000\AA\ break region (right panel). From the highly
star-forming SF1 class to the red-sequence we see the expected steady
decrease in \oii\ emission line strength, decrease in blue/UV flux,
increase in 4000\AA\ break strength, and strengthening of the Ca\,H\&K
absorption lines, all indicating that specific star formation rate
decreases and mean stellar age increases as we move through these
classes. Weak \oii\ emission is visible in the red-sequence stack,
which may indicate the presence of a narrow line (obscured) AGN in
some objects. The \neiii $\lambda$3869\AA\ emission line is
visible in the SF1 class, consistent with the high ionisation state of
gas in young starbursts.

\begin{figure}
\hspace*{-0.25cm}
\includegraphics[scale=0.9]{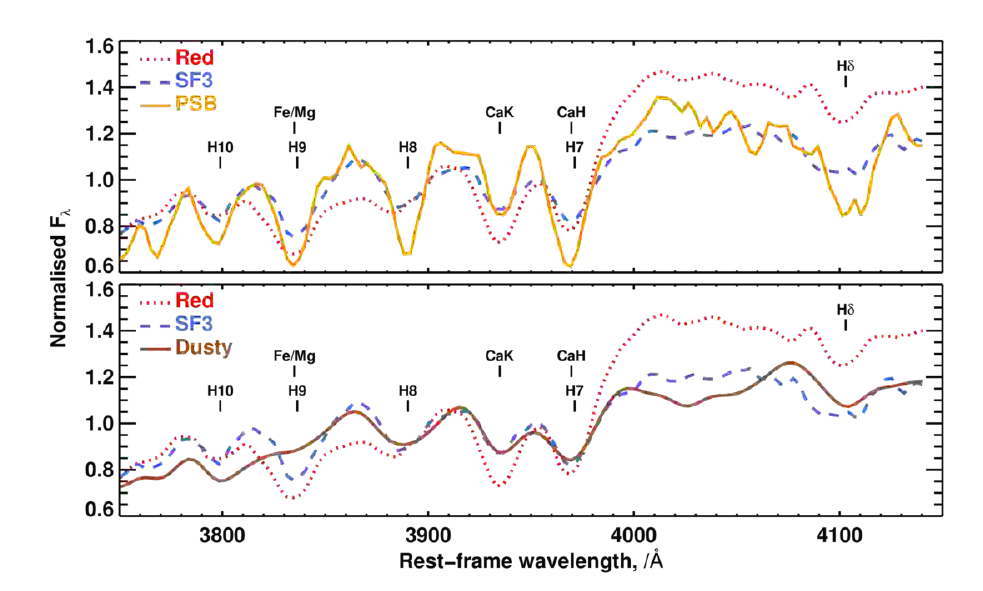}
\caption{Comparison of the 4000\AA\ region of the post-starburst and
  dusty samples with the red-sequence and SF3 (oldest blue-sequence)
  classes, with the main absorption features identified. This
  highlights the strong Balmer lines of the post-starburst population,
  and the close match of the dusty-starforming galaxy spectrum to an
  ordinary blue-cloud galaxy.}\label{fig:stackcompare}
\end{figure}

In Figure \ref{fig:stackcompare} we compare the 4000\AA\ break region
of the more unusual dusty and post-starburst classes with the
red-sequence and SF3 classes, which have the most similar SED
shapes. The top panel clearly shows the strong Balmer absorption lines
of the post-starburst galaxies compared to the SF3 class, and the
different continuum shape compared to the red-sequence galaxies caused
by the post-starburst galaxies having a Balmer break rather than
4000\AA\ break. The bottom panel shows that the dusty star-forming
galaxies have absorption line and break strengths that are very similar to the
SF3 class. In the left panel of Figure \ref{fig:stacks} we also see
that they have a similar strength of \oii\ emission line, although the
dusty star-forming galaxies have a noticeably redder UV continuum
slope, as expected from their super-colour classification.

\subsubsection{Selecting post-starburst galaxies}

129 galaxies have spectra with sufficient SNR to measure their spectral
indices using the method developed by \citet[][hereafter
WWJ09]{Wild:2009p2609} for galaxies in the VIMOS VLT Deep Survey
(VVDS, see Appendix \ref{app:vvds}). This allows us to position the
super-colour demarcation line between post-starburst and red-sequence
galaxies such that all the spectroscopically identified post-starburst
galaxies (4) lie in the post-starburst super-colour class.  Galaxies
with lower SC1 and SC2 values have optical spectra that were consistent
with being in the red-sequence as defined by the VVDS spectral
indices. One galaxy has spectral indices placing it in between the
red-sequence and post-starburst classes, and this galaxy also lies to
the red end of the post-starburst class in super-colour space. We set
the boundary to include this object, although this means that our
photometrically selected post-starbursts may extend to slightly older
ages than the spectroscopically selected post-starbursts in WWJ09.

%------------------------------------------------------------------
\subsection{Comparison to traditional colour-colour diagrams}\label{sec:uvj}
%------------------------------------------------------------------

\begin{figure*}
\includegraphics[scale=1]{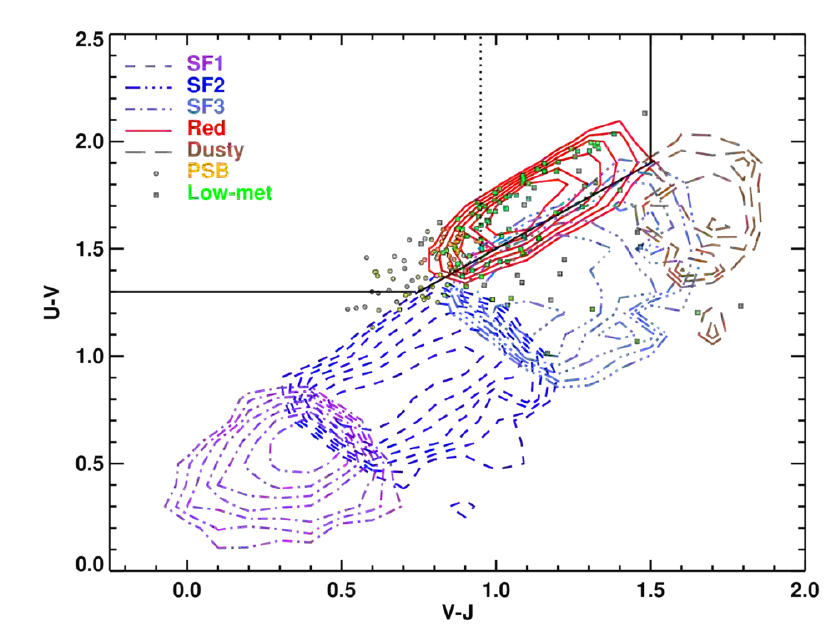}
\includegraphics[scale=1]{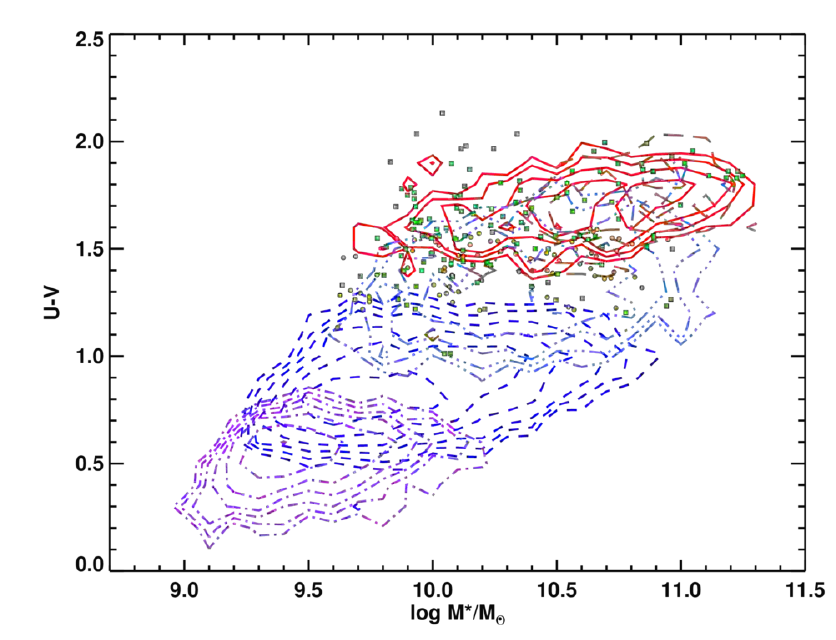}
\caption{The K-corrected rest-frame $UVJ$ colour-colour diagram (left)
  and colour-stellar mass diagram (right) of the UDS galaxies with
  $0.9<z<1.2$. Contours show the loci of the primary classes as
  determined from their super-colours, symbols show where galaxies
  belonging to the two rarer classes lie. Black lines in the left hand
  panel indicate the
  standard demarcation lines between red-sequence, blue-sequence
  and dusty-starforming galaxies. The dotted black line shows the cut used by
  \citet{Whitaker:2012p8738} to separate post-starburst galaxies from
  red-sequence galaxies. }\label{fig:uvj}
\end{figure*}

In recent years, the rest-frame (K-corrected) UVJ colour-colour
diagram has been the preferred method to separate red quiescent from
blue-starforming and dusty-starforming galaxies
\citep[e.g.][]{Wuyts:2007p8937,Williams:2009p8926,Ilbert:2013p9113},
and to identify ``young red-sequence'' or post-starburst
galaxies \citep{Whitaker:2012p8738}. It is therefore valuable to
compare the PCA derived super-colours with the UVJ colours. For each
galaxy in our sample we perform a traditional K-correction by
identifying the best-fit stochastic burst model SED, and then
measuring the rest-frame U-V and V-J colours from the best-fit
model\footnote{We note that some authors extrapolate from the model
  colours in an attempt to allow for poor model fits (e.g. Williams et
  al. 2009, Whitaker et al. 2011, following Rudnick et
  al. 2003)\nocite{Williams:2009p8926, Whitaker:2011p8798,
    Rudnick:2003p8956}. We do not attempt to reproduce this method
  here and it is not simple to assess the success of the extrapolation
  as a function of redshift, SED type and assumed model
  templates. }. To reproduce previous results from the literature we
use the \citet{Bessell:1990p8929} $U$ and $V$ filters, and the UKIRT
$J$ filter which is on the Mauna Kea system
\citep{Tokunaga:2002p8948}. The resulting UVJ diagram is shown in the
left panel of Figure \ref{fig:uvj}, with objects colour coded by the
super-colour SED classification described in the previous subsection.

It is not a surprise to see that the two methods produce very
consistent classifications between the major classes: as with
optical spectra, PCA has identified the major variations of galaxy
SEDs to be exactly those that astronomers have identified by eye. The
advantage of the PCA is the combination of many filters into single
colours to improve robustness and signal-to-noise, the independence
from model fitting, and the ability to identify higher order
variations as seen in the third component.

There are some details of this comparison that are worth
highlighting. Firstly, for blue-cloud galaxies, the cuts of constant
SC1 slice across the UVJ colour distribution almost perpendicular to
the axis of the blue-cloud in UVJ colour space. This illustrates how
PCA rotates colour-colour space to combine colours which are
correlated (i.e. colours that provide equivalent 
information). Secondly, the post-starburst population are found at the
bluest end of the red-sequence in the rest-frame UVJ colour-colour
diagram, as suggested by \citet{Whitaker:2012p8738},  however their
  cut at $V-J<0.95$ includes a majority of objects that we are unable to confirm as
  post-starburst galaxies based on our comparison to the UDSz
  spectra. Low-metallicity red-sequence galaxies scatter throughout
  the red sequence and SF3 classes; this colour combination does not
  allow the separation of this class. Finally, the very tight
red sequence identified by the PCA is considerably more extended in
UVJ colour space, and has significant overlap with the low-level star-forming
galaxies. 

In the right hand panel of Figure \ref{fig:uvj} we show the
colour-stellar mass diagram of the galaxies in our catalogue. Stellar
masses are calculated from the stellar mass-to-light ratio of the
best-fit spectral synthesis model and the flux normalisation at
1\mum\ output from the PCA. When the V-J colour is not considered, the
red-sequence is clearly contaminated with dusty star-forming galaxies
and older/dustier/more metal rich star-forming galaxies. The
so-called ``green-valley'' between blue and red-sequences is comprised
of predominantly older/dustier/more metal rich star-forming  galaxies, but also contains a large
fraction of the post-starburst galaxies \citep[see
also][]{Wong:2012p8892}. The low-metallicity class lies predominantly
at the low-mass end of the red-sequence.  We will compare the stellar
masses functions of the different classes in Section \ref{sec:LFs}.

%%%%%%%%%%%%%%%%%%%%%%%%%%%%%%%%%%%%%%%%%%%%%%%%%%%%%%%%%%%%%%%%%%
\section{Results}\label{sec:results}
%%%%%%%%%%%%%%%%%%%%%%%%%%%%%%%%%%%%%%%%%%%%%%%%%%%%%%%%%%%%%%%%%%
In the previous sections we presented a method to define linear
combinations of filters that optimally describe the shapes of galaxy
SEDs. We applied this method to galaxies in the SXDS/UDS field, and
qualitatively confirmed our interpretation of the physical properties
of each class by stacking rest-frame optical spectra. In this Section
we stack the SEDs of the galaxies in each class, and fit spectral
synthesis models to obtain quantitative physical properties. We then
construct luminosity and mass functions of each class.

%------------------------------------------------------------------
\subsection{Stacked SEDs and average physical properties}
%------------------------------------------------------------------

\begin{figure*}
\includegraphics[scale=1]{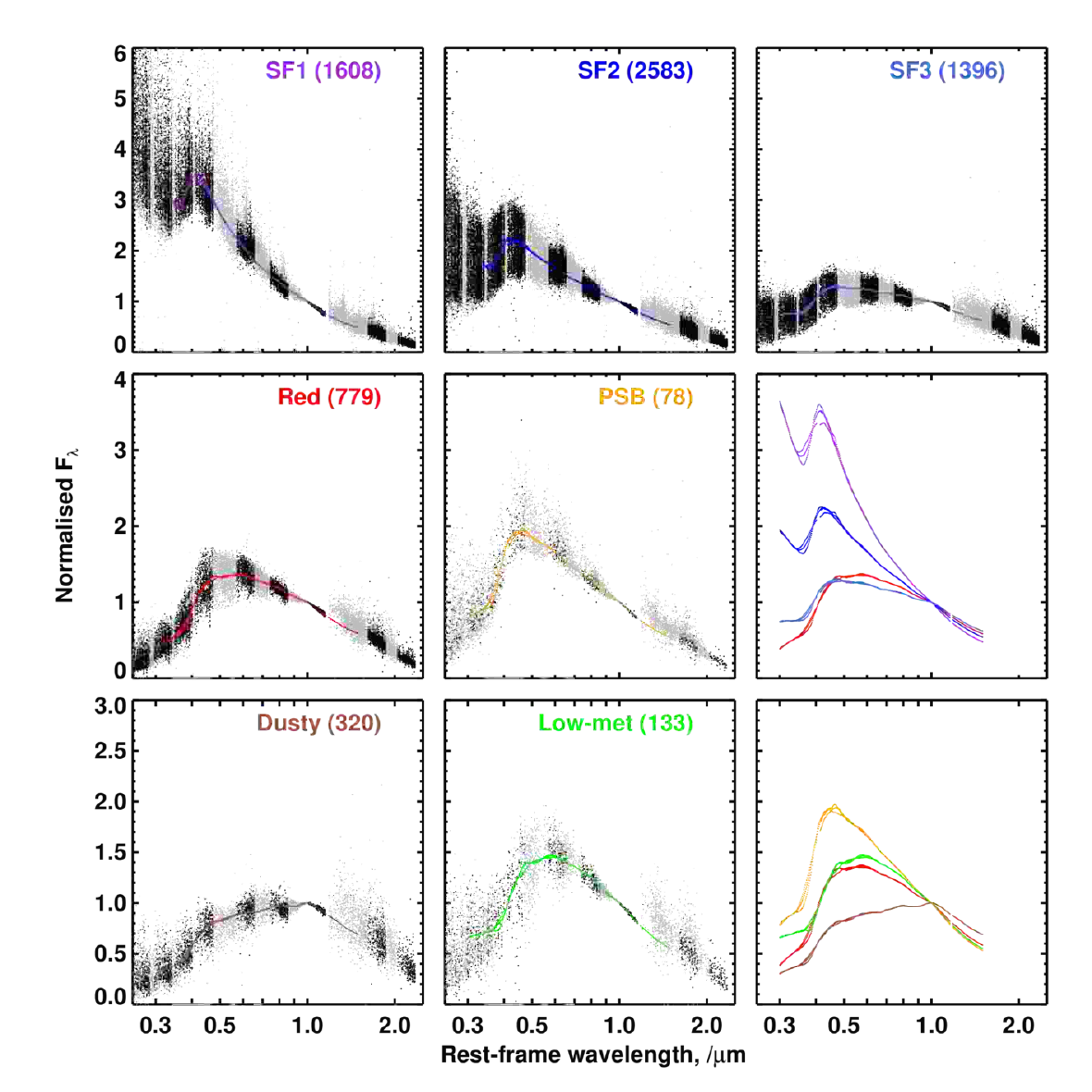}
\caption{Combined spectral energy distributions for each super-colour
  defined galaxy class, created from the individual photometry of all
  galaxies in the class, normalised and shifted to the
  rest-frame. From left to right and top to bottom: the three
  star-forming classes (SF1 to SF3), red-sequence, post-starburst,
  dusty star-forming and low-metallicity quiescent galaxies. The black
  (grey) points indicate fluxes contributed by galaxies in the low-
  (high-) redshift sample split at $z=1.2$. In this paper we analyse
  only the low-redshift sample where spectra are available and the
  errors on the super-colours are smaller, the high-redshift sample is
  included in this figure only for illustration. A much larger scatter
  in the combined SEDs is apparent in the high redshift sample,
  reflecting the increased errors on the SCs and photometric
  redshifts.  The
  number of $0.9<z<1.2$ galaxies in each sample is given in brackets
  in each panel. The PCA reconstruction
  of the median SED of each class is overplotted as a coloured line. The
  super-colours corresponding to these reconstructions are given in
  Table \ref{tab:averageSC}.  The central and lower right panels
  overplot the PCA reconstructed SEDs for some of the classes, to
  allow easy comparison between the SED shapes.}\label{fig:seds}
\end{figure*}

 \begin{table}
   \begin{center}
     \caption{Average super-colours of each galaxy class, derived
       from the fits to the combined SEDs for $0.9<z<1.2$ galaxies
       presented in Figure \ref{fig:seds}. The second column gives the
     number of contributing galaxies. }
\label{tab:averageSC}
     \begin{tabular}{ccccc}\hline\hline
       Class & Number & SC1 & SC2 & SC3 \\ \hline
       SF1 & 1608 & 30.2$\pm$0.9 & -1.0$\pm$0.9 & 0.9$\pm$0.5 \\
       SF2 & 2583 & 4.0$\pm$1.0 & -1.2$\pm$0.9 & 0.5$\pm$0.5 \\
       SF3 & 1396 & -16.9$\pm$0.5 & -3.6$\pm$0.4 & -0.4$\pm$0.3 \\
       Red & 779 & -19.3$\pm$0.3 & -0.4$\pm$0.3 & 0.8$\pm$0.2 \\
       PSB & 78 & -7.1$\pm$0.5 & 4.2$\pm$0.5 & 0.3$\pm$0.5 \\
     Dusty & 320 & -27.3$\pm$0.2 & -7.2$\pm$0.2 & -1.9$\pm$0.2 \\
   Low-met & 133 & -17.0$\pm$0.4 & -0.9$\pm$0.4 & 2.9$\pm$0.3 \\

       \hline
     \end{tabular}
   \end{center}
%   Col. (1): Class name\\
 %  Col. (2): Number of objects\\
  % Col. (3-5): Super-colours of the median SED \\
 \end{table}

The UDS galaxies cover a wide range of redshifts, therefore while a
single galaxy is observed at only $\sim$6 sparsely sampled wavelength
points, the full wavelength range is observed within the sample. In
Figure \ref{fig:seds} we exploit this to show the average SEDs of each
class. The number of $0.9<z<1.2$ galaxies in each class is shown in
each panel in brackets. This figure highlights the distinctly
different SED shapes that have been identified by the PCA. We show the
$0.9<z<1.2$ sample as black points: the smaller errors on both the
photometric redshifts and on the super-colours compared to higher
redshifts are evident in the much reduced scatter in the combined
SEDs, at a given wavelength. We include the higher redshift objects in this
plot to help visualise the full SED and note features such as the \ha\
emission line visible in the starburst galaxies.
 
We calculate the typical super-colours of each class from
the median normalised flux of all ($0.9<z<1.2$) galaxies which contribute to each
wavelength bin, and take the 16th and 84th percentiles as estimates of the
errors on these fluxes. The resulting typical super-colours of each class are
given in Table \ref{tab:averageSC} and the reconstructed SEDs are
overplotted in each panel of Figure \ref{fig:seds}.

\begin{table*}
  \begin{center}
    \caption{Typical physical properties of each galaxy class, estimated from
      a library of 440,000 BC03 spectral synthesis models. For each
      parameter the 16th, 50th and 84th percentiles of the probability
      distribution function are given. Note that for the dusty and
      low-metallicity samples the high reduced $\chi_\nu^2$ value indicates that
      the models are unable to fit the shape of the SED and the
      derived parameters and errors should be treated with
      caution.}\label{tab:averageSED}
    \begin{tabular}{ccccccccc}\hline\hline
      Class & $\chi^2_\nu$ & age$_{\rm wr}$/Gyr & age$_{\rm wm}$/Gyr & Z/Z$_\odot$ & $\tau_V$ &F$_{\rm burst}$ & log(M/L)$_V$ & log(M/L)$_K$ \\ \hline
       SF1 & 0.7& 0.4,0.4,0.4& 0.8,0.8,1.0& 0.8,1.2,1.3& 0.2,0.2,0.5& 0.0,0.0,0.1& -0.61,-0.61,-0.55& -0.90,-0.89,-0.80 \\
       SF2 & 0.1& 0.7,1.0,1.3& 1.3,1.6,2.5& 1.2,1.5,1.7& 0.2,0.5,0.8& 0.0,0.0,0.0& -0.34,-0.28,-0.22& -0.79,-0.70,-0.64 \\
       SF3 & 0.3& 1.6,2.1,2.4& 2.2,3.1,3.6& 1.6,1.7,1.9& 0.8,1.1,1.7& 0.0,0.0,0.0& 0.12,0.15,0.21& -0.57,-0.52,-0.48 \\
       Red & 0.2& 1.3,1.8,2.4& 1.6,2.2,3.1& 0.5,1.0,1.5& 1.1,1.4,2.4& 0.0,0.0,0.1& 0.03,0.12,0.21& -0.59,-0.50,-0.40 \\
       PSB & 0.2& 0.7,1.0,1.6& 0.8,1.3,2.5& 0.5,1.0,1.6& 0.5,1.4,2.4& 0.2,0.4,0.7& -0.25,-0.16,-0.10& -0.74,-0.62,-0.52 \\
     Dusty & 1.6& 0.7,1.0,1.6& 1.3,1.6,2.5& 1.5,1.9,1.9& 2.1,2.7,4.9& 0.0,0.0,0.0& 0.57,0.60,0.66& -0.42,-0.42,-0.38 \\
   Low-met & 17.8& 2.7,3.2,3.8& 3.1,3.6,4.5& 0.4,0.6,0.8& 0.2,0.8,1.7& 0.0,0.0,0.0& 0.09,0.18,0.24& -0.43,-0.37,-0.30 \\

      \hline
    \end{tabular}
  \end{center}
\begin{flushleft}
  Col. (1): Class name\\
  Col. (2): Reduced $\chi_\nu^2$ of the best-fit BC03 stochastic
  burst model \\
  Col. (3-9): 16th, 50th and 84th percentiles of the probability
  distribution function of physical (derived) parameters: $r$-band
  light-weighted age; mass-weighted age; metallicity; total effective
  $V$-band optical depth due to dust attenuation; fraction of stars
  formed in a starburst in the last Gyr; log mass-to-light ratio in
  the $V$ and $K$ bands (we assume solar AB absolute magnitudes of 4.8
  and 5.18 in the $V$ and $K$ bands respectively).
 \end{flushleft}
\end{table*}

\begin{figure*}
\includegraphics[scale=1]{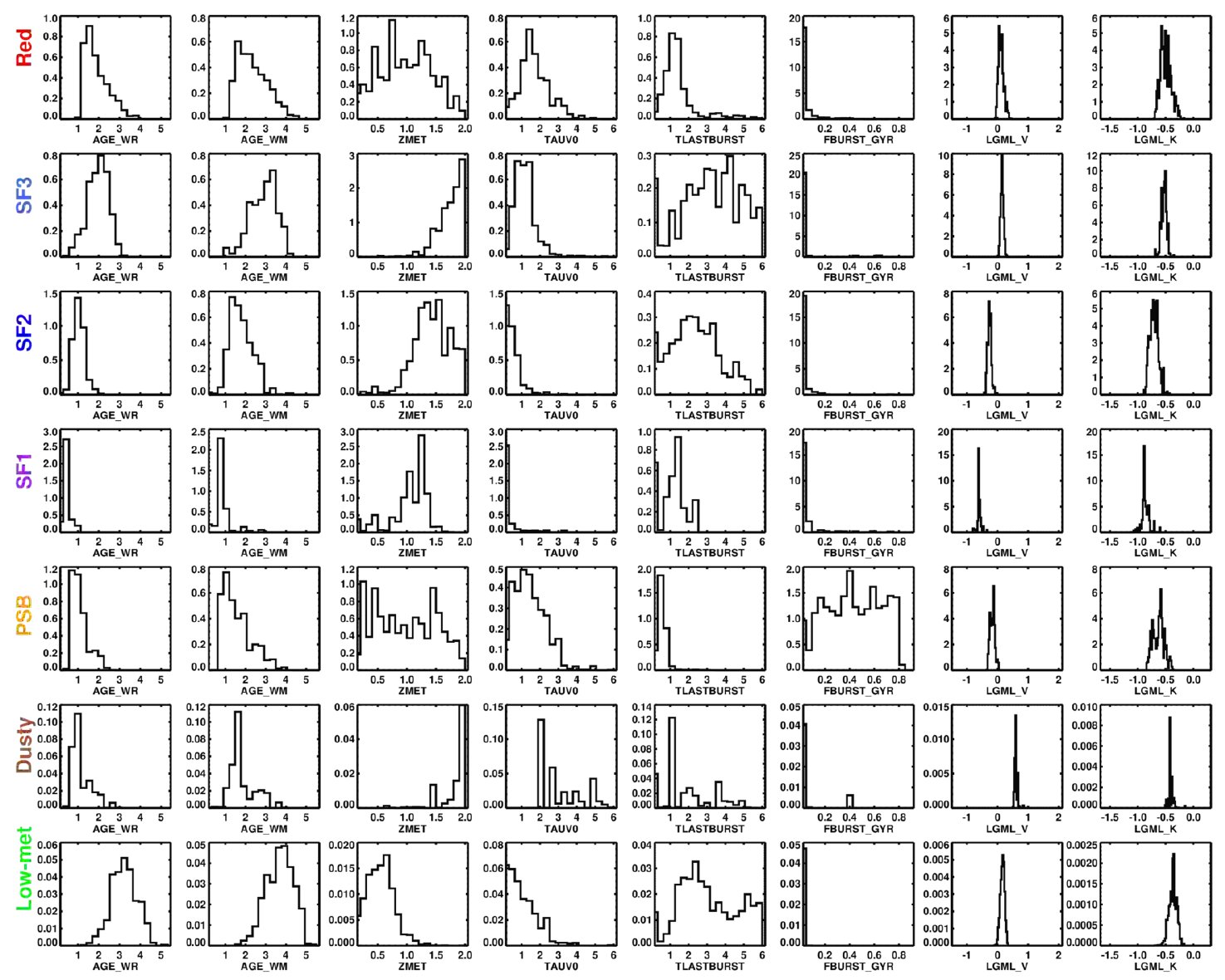}
\caption{Probability distribution functions of a selection of physical
  parameters for each of the super-colour classes. From left to right:
  $r$-band light-weighted age (Gyr); mass-weighted age (Gyr);
  metallicity (Z/Z$_\odot$); total effective $V$-band absorption
  optical depth of the dust seen by young stars inside birth clouds;
  time of the last burst of star-formation (Gyr); fraction of stars
  formed in a starburst in the last Gyr; log mass-to-light ratio in
  the $V$-band and in the $K$-band.  }\label{fig:pdfs}
\end{figure*}

For each of these typical SEDs we calculate the probability
distribution function (PDF) of derived (model) parameters, using the
stochastic burst models as a prior. The method is similar to that
described in detail in e.g.  \citet{2003MNRAS.341...33K} and
\citet{Gallazzi:2005p6450} for fitting models to SDSS data using
spectral indices. The 16th, 50th and 84th percentiles of the PDFs are
given in Table \ref{tab:averageSED} for a selection of parameters, and
the full PDFs are shown in Figure \ref{fig:pdfs}.  The reduced
$\chi_\nu^2$ of the best-fit model indicates the ability, or
otherwise, of the models to fit the observed SED shape. In the case of
the dusty galaxies and low-metallicity galaxies, this value is greater
than unity, meaning the models provide a poor fit to these classes. Therefore the quoted physical parameters should
be taken as indicative only\footnote{Because of the poor model fits
  and small errors on the data afforded by the stacking of many
  galaxies, the PDFs for the dusty and low-metallicity classes are
  dominated by a single ``closest'' model spectrum. When estimating
  the physical parameters of these two classes, we therefore increased
  the errors on their super-colours by a factor of 1.5. This allowed a
  sufficient number of models to lie within the error range to obtain
  an estimate of the typical parameters of a range of the closest
  matching models. The precise factor chosen makes no significant difference to
  the results, which should be taken as indicative only. }.

In agreement with the trends seen in the stacked optical spectra, we
find an overall increase in mean stellar age from SF1 to SF3
galaxies. In contrast to what is expected from the optical spectra,
however, the SF3 class is fit with models of an older mean stellar age
than the red-sequence class. The SF3 class is fitted with models with
a slower exponential decay time for star formation than the
red-sequence class class (3.3\,Gyr compared to 1.4\,Gyr), and an
earlier time of formation. To obtain more robust ages, there is
clearly a need for a careful analysis of the degeneracies between age,
dust and metallicity. Given the limitations of our SED fitting with
respect to the fixed dust attenuation law, fixed metallicities,
parameterised star formation histories and need to account for the
variations between spectral synthesis models, we do not interpret
these trends further.

Turning to the post-starburst, low-metallicity and dusty classes, the
super-colours and errors show that these have statistically different
SED shapes compared to the more usual classes, at high significance
(Table \ref{tab:averageSC}). The post-starburst class is clearly
identified from the SED fit as the only class with a significant
recent burst mass fraction. The fraction of stars formed in a burst in
the last 1\,Gyr is not well constrained by the super-colours, but is
larger than $\sim$10\%. While the post-starburst class is well fit by
the models, the conclusions we can draw about the physical properties
of the dusty and low-metallicity classes are based on extrapolation
from the closest models.  The dusty galaxies are fit with models with
total effective $V$-band absorption optical depth of the dust seen by
young stars inside birth clouds of $\tau_V\sim2$, which is
significantly higher than all other classes. Their fitted
mean stellar age is similar to the blue-cloud galaxies (SF2), although
the stacked optical spectra show a closer match to the SF3 class. We
note that the old, metal-rich nature of this population is almost
certainly a selection effect: only the oldest, most metal rich dusty
star-forming galaxies will be uniquely identifiable at the very
reddest end of the blue-sequence. Younger and less metal rich dusty
galaxies will be indistinguishable from normal blue-sequence galaxies
with broad band photometry alone. The low-metallicity class is fit
with models with a metallicity of less than solar. Interestingly, the
mean stellar ages of this population are the oldest of all
classes. This suggests that we are able to select the oldest, most metal
poor quiescent galaxies at a given redshift from SED shape alone.

%------------------------------------------------------------------
\subsection{Luminosity and Mass Functions}\label{sec:LFs}
%------------------------------------------------------------------

\begin{figure*}
\includegraphics[scale=1]{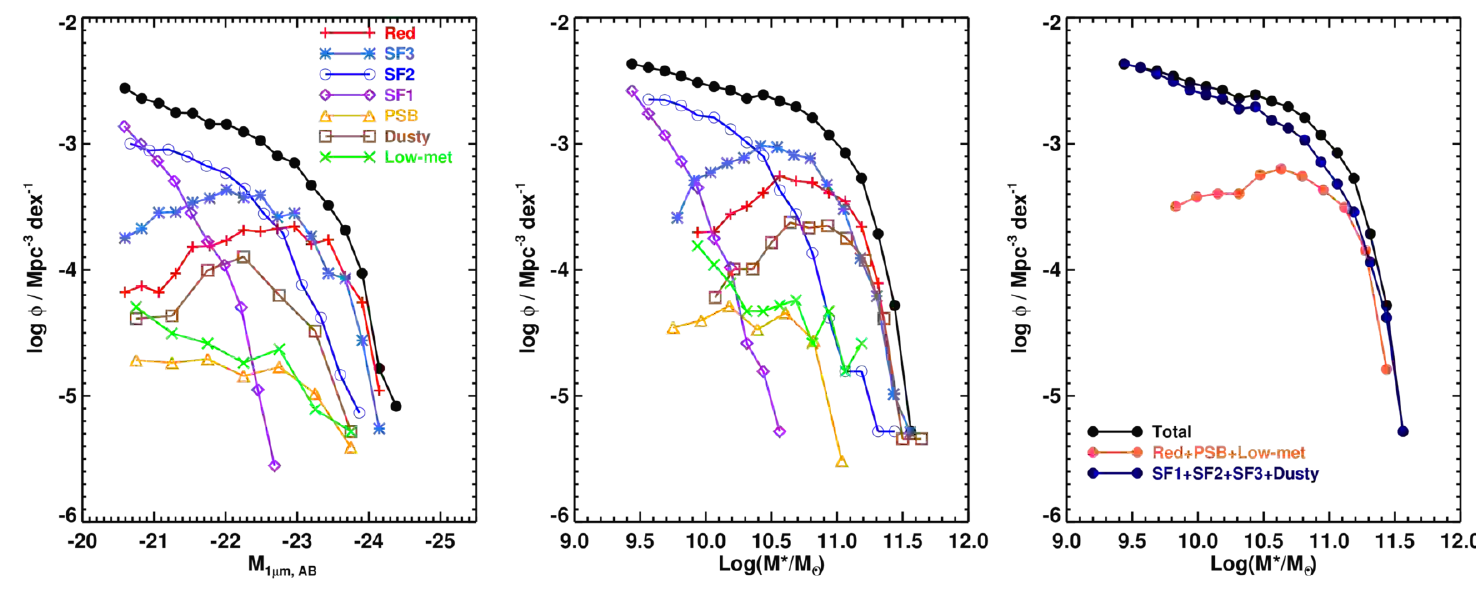}
\caption{\emph{Left:} The 1\mum\ luminosity functions of the complete
  sample (black) and individual classes (colours). For comparison to
  the literature the extrapolated K-band luminosity function is shown in
  Appendix \ref{app:KLF}. \emph{Centre:} Stellar mass functions of the
  complete sample (black) and individual classes
  (colours). \emph{Right: } Stellar mass functions of the complete
  sample (black), red-sequence (comprising the red, PSB and low-metallicity
  classes) and blue-cloud (comprising the SF1, SF2, SF3 and dusty
  classes).  Stellar mass is defined as the current observed mass of
  stars in the galaxy, i.e. not including mass returned to the ISM
  through winds and supernovae. }\label{fig:lfs}
\end{figure*}

In Figure \ref{fig:lfs} we present the rest-frame 1\mum\ luminosity
functions and stellar mass functions of the different galaxy classes,
accounting for volume effects using the V$_{\rm max}$ method. 
  This weights each galaxy by 1/V$_{\rm max}$, where V$_{\rm max}$ is
  the maximum volume in which the galaxy may be observed in the
  spectroscopic survey \citep{1968ApJ...151..393S}. We use the
  best-fit stochastic burst model for each galaxy in order to
  calculate how the apparent magnitude changes as a function of
  redshift, and therefore the volume in which a galaxy of that SED
  shape and absolute magnitude would be seen.  To avoid being
dominated by incompleteness correction, we only show bins where the
median visible volume of galaxies in the bin is greater than 60\% of
the total survey volume. It is important to recall that the spectral
synthesis models do not provide a good fit to the dusty star-forming
galaxies, low-metallicity galaxies and some of the extreme starbursts
and therefore the mass-to-light ratios have greater uncertainties than
for other classes, even though for individual galaxies the fit is
formally acceptable. This may result in stellar masses and volume
corrections that are biased in an unknown way due to the failure of
the models to match their colours.  We have chosen to present
luminosity functions at 1\mum\ as this is the normalisation point of
the PCA, to avoid model dependent extrapolation. $K$-band luminosity
functions are additionally presented in Appendix \ref{app:KLF} to
facilitate direct comparison with other results in the literature
\citep[e.g.][]{Cirasuolo:2010p8794}.

Three populations dominate at high stellar masses in almost equal
proportions ($\log{\rm M^*}>11$): the red-sequence, high mean stellar
age blue-cloud (SF3) and dusty star-forming classes. Within the
blue-cloud we see the well known trend between mean stellar age and
stellar mass. The ``normal'' blue-cloud class (SF2) dominates the
population at masses below $\log{\rm M^*}\sim10.5$, and galaxies in
the youngest class (SF1) have predominantly low masses ($\log{\rm
  M^*}<10$). Turning to the unusual classes, the mass distribution of
the dusty-starforming galaxies closely matches that of the
red-sequence galaxies, and is different from any of the other star
forming classes. The post-starburst galaxies have predominantly
intermediate masses ($\log{\rm M^*}<10.75$) and their mass distribution
is uniquely rather flat.  The low-metallicity red-sequence galaxies
have a stellar mass function that climbs steadily towards low mass
($\log{\rm M^*}<10$).

In the right panel we combine all of the quiescent and star-forming
classes into two independent luminosity functions, to compare with
similar analyses in the literature. We will return to this figure in
Section \ref{sec:bimodal} below.

%%%%%%%%%%%%%%%%%%%%%%%%%%%%%%%%%%%%%%%%%%%%%%%%%%%%%%%%%%%%%%%%%%
\section{Discussion}\label{sec:disc}
%%%%%%%%%%%%%%%%%%%%%%%%%%%%%%%%%%%%%%%%%%%%%%%%%%%%%%%%%%%%%%%%%%
\begin{figure*}
\includegraphics[scale=1]{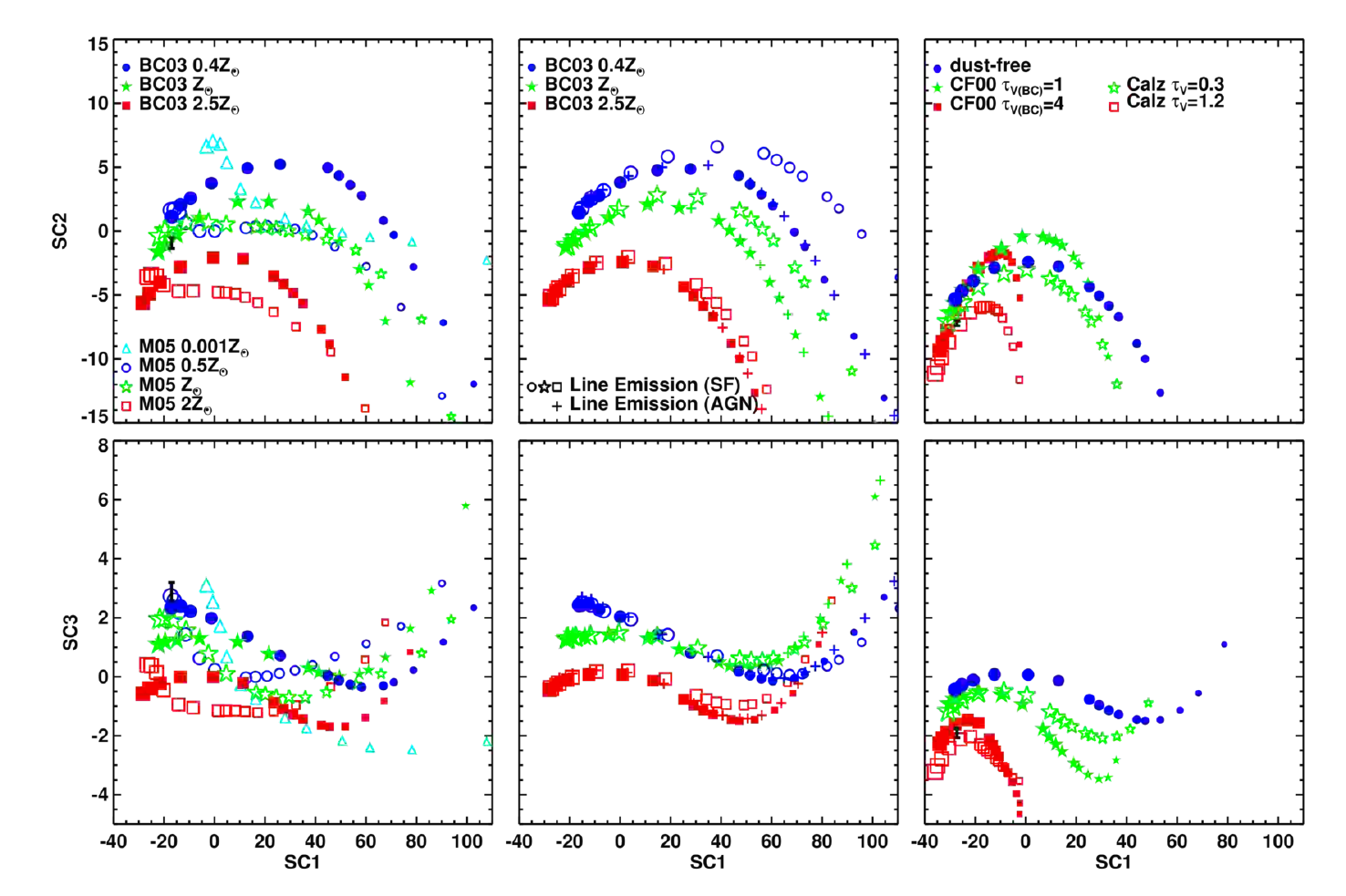}
\caption{Comparison of super-colours of different stellar population
  models. All panels show the evolution of colours of an exponentially declining
  star formation history with an e-folding time of 1\,Gyr; the symbol
  size increases in size with increasing age from 100\,Myr to 6\,Gyr
  in bins of 100\,Myr until 1Gyr and then bins of 1\,Gyr until 6\,Gyr,
  with an additional bin at 1.5\,Gyr. \emph{Left:} BC03 and
  \citet{Maraston:2005p8966} models, both with Salpeter IMF, with
  black error bar indicating the mean colour of the low-metallicity
  class of galaxies; \emph{Centre:} BC03 models with Chabrier IMF,
  with and without nebular emission lines; \emph{Right:} two different dust
  attenuation laws applied to the twice solar metallicity BC03 models,
  with the black error bar indicating the mean colour of the dusty
  class of galaxies. }
\label{fig:M05}
\end{figure*}

Previous analyses of galaxy SED shapes in broad-band photometric
surveys can be split into two categories: SED-fitting and rest-frame
colour-colour diagrams. In this paper we have presented a third
alternative which allows a fresh look at both interesting populations
of galaxies and the ability of spectral synthesis models to fit all
SED types.

In this section we discuss the impact that different stellar
population models, inclusion of emission lines and changing the dust
attenuation law would have on our results and the implications for
future improvement of spectral synthesis models and SED-fitting
codes. We then summarise the implications of our results for three
science topics that could be investigated further using this method.

%------------------------------------------------------------------
\subsection{BC03 vs. Maraston models}\label{sec:M05}
%------------------------------------------------------------------
One key difference between some stellar population models is the relative
strength of emission from the asymptotic-giant branch (AGB) phase of
stellar evolution, which primarily alters the optical-to-NIR flux
ratio of galaxies with stars younger than 2\,Gyr. Attempts to
positively identify AGB features in the NIR of integrated stellar
populations have had mixed results [see Kriek et~al. (2010) and Zibetti
et~al. (2013), compared to Melbourne et al. (2012) and Martins
et~al. (2013)]\nocite{Kriek:2010p8971,Melbourne:2012p9128,Zibetti:2013p9091,Martins:2013p9118}.

In agreement with \citet{Kriek:2010p8971} and
\citet{Zibetti:2013p9091}, we see no evidence for a deviation of
colours from those predicted by the BC03 models in post-starburst
galaxies (Figure \ref{fig:seds} and Table \ref{tab:averageSED}),
although our SED analysis does not extend as far into the NIR as these
two studies. The positive identification of AGB features by both
Melbourne et~al. (2012) and Martins et~al. (2013) are in galaxies with
younger mean stellar ages, i.e. star-forming and starburst galaxies.
 The super-colours can be used to investigate the potential impact of
different assumptions about this late stellar evolutionary phase on
the SED shapes of galaxies.  

In the left panels of Figure \ref{fig:M05} we compare the
super-colours of the AGB-light BC03 stellar population synthesis
models with the AGB-heavy models of \citet[][hereafter
M05]{Maraston:2005p8966}. For the purposes of this investigation we
focus on simple models with exponentially declining star formation
histories with an e-folding time of 1\,Gyr, and no dust attenuation. A range
of metallicities are shown.  As the stellar population ages the overall
blue-to-red colour (SC1) does not vary significantly between the two
models, but the same is not true for the features measured by SC2 and
SC3 (the shape of the 4000\AA\ break region). The largest difference
occurs in the normal star-forming galaxies, particularly at sub- and
super-solar metallicities. Unfortunately, this region has the least
well constrained stellar population balance, and it will be difficult to
break the degeneracy between TP-AGB star contribution and star
formation history from photometry alone.

The lower panel shows how SC3 increases as metallicity decreases in
the red-sequence. The black error bar indicates the mean colour of the
low-metallicity class (the error on SC1 is insignificant on this
scale).  In contrast to the BC03 models, the M05 models reach a high
enough value of SC3 to fit the low-metallicity class, although the
reason for this difference is unclear.  Further testing with realistic
star formation histories would be required to see whether M05 models
can provide an improved overall fit to this class than the BC03
models, for which a reduced $\chi_\nu^2$ of 17.8 was obtained.

While the BC03 models used in this paper extend down to 1/10th solar,
the extreme low metallicity (1/1000th solar) M05 model included in
this figure highlights another area where careful interpretation of
the SED shapes is needed: at ages larger than $\sim$4\,Gyr extreme
low-metallicity galaxies have SC1/2 colours consistent with a
post-starburst population. While the UVJ post-starburst selection
method may include low-metallicity quiescent galaxies, the third
super-colour allows us to cleanly separate these two interesting
populations.

%------------------------------------------------------------------
\subsection{The impact of nebular emission lines}\label{sec:emlines}
%------------------------------------------------------------------
Narrow emission lines and blue continuum emission can arise from
either HII regions in sites of active star formation or from obscured
AGN. Their omission during SED fitting has been shown to bias the
physical properties of galaxies with very high specific star formation
rates \citep[e.g.][]{Schaerer:2009p4168}. While nebular continuum
emission is not expected to be strong in the optical and NIR
wavelength range studied here, strong nebular emission lines may be
present. In the central panels of Figure \ref{fig:M05} we investigate
the impact of line emission on the super-colours, by adding the
strongest optical emission lines to the spectra before calculating the
observed frame broad band fluxes.  The model emission lines were not
attenuated by dust, thus giving the maximum possible impact. 

 The open circles show the impact on the super-colours due to emission
from star formation regions. The \ha\ line strength was estimated from
the number of ionising photons from the UV stellar continuum below
912\AA\ and the hydrogen Balmer series were included up to \he\
assuming Case B recombination line ratios
\citep{Osterbrock:2006p4475}. Strong metal lines were included using
dust attenuation corrected line ratios observed for local star-forming
galaxies of appropriate metallicity from \citet{Pacifici:2012p9111}
(\oiii$\lambda\lambda4960,5008$, \oii$\lambda\lambda3727, 3730$,
\nii$\lambda\lambda6550,6585$, \sii$\lambda\lambda6718,6733$).  We
observe a small increase in SC1 and SC2 in young galaxies when
emission lines from star formation are included. The shift is most
significant in SC2 for very young, low metallicity galaxies. Only a
small effect is observed on SC3.

The crosses in the central panel of  Figure \ref{fig:M05} show the
effect of emission from obscured (narrow-line) AGN by
including a fixed line luminosity independent of the age of the
stellar population. We assume a fiducial AGN
\oiii\ luminosity of $10^7$L$_\odot$ \citep{heckman04} and include
strong metal lines with line ratios typical for narrow line AGN in the
local Universe \citep{2006MNRAS.372..961K}. The impact of these lines
is most significant on the intermediate age population, causing a
small shift in SC1.

Comparing to Figure \ref{fig:pc123} emission lines are not able to
explain the high SC3 values of some of the youngest starbursts, nor
are they able to explain the high SC3 values of the low-metallicity
class of galaxies (black error bar in Figure \ref{fig:M05}). However,
it is possible that the impact of emission lines is greater at certain
redshifts, when the positioning of the filters happens to coincide
with the strongest lines. The super-colours provide a new method for
investigating this effect in much more detail in the future.

%------------------------------------------------------------------
\subsection{The impact of the assumed dust attenuation law}\label{sec:dust}
%------------------------------------------------------------------
 In the right hand panel of Figure \ref{fig:M05} we investigate the
sensitivity of the super-colours to the assumed shape of the dust
attenuation law. We compare the colours of the exponentially declining
model attenuated using the \citet[][CF00]{2000ApJ...539..718C}
prescription employed in the stochastic burst models and by the
\citet{Calzetti:2000p4473} attenuation law. This latter attenuation
law lies at the grey end of the range of attenuation laws found in
local star-forming galaxies \citep{Wild:2011p6890}, but is generally
used for high-redshift galaxies. Note that the CF00 prescription is
parameterised by the effective optical depth in the $V$-band towards
stars younger than $10^7$ years, which is independent of mean stellar
age of the population. On the other hand, the Calzetti prescription is
parameterised by the effective optical depth in the $V$-band towards
all stars. This makes it difficult to directly compare the evolution
in colours for a fixed dust content over a wide age range. For old
stellar populations, the Calzetti value is a factor of $\sim$3 lower
than the CF00 value for the same amount of dust. As the
dusty-starforming class has an old mean stellar age, we normalise the
attenuation laws to best compare the oldest populations. At younger
ages, the Calzetti attenuated models will have lower effective
attenuation than the CF00 attenuated models.

In the upper right panel of Figure \ref{fig:M05} we can clearly see
the dust-age degeneracy, with dustier star-forming galaxies moving to
lower SC1 values at all ages. Interestingly, the Calzetti attenuation
law has a much larger impact on the SC2 values than the CF00
attenuation law. In the lower panel, we see the dust-metallicity
degeneracy, with dustier galaxies moving to lower SC3 at all
ages. 

In Section \ref{sec:results} we noted that the mean SED of the dusty
star-forming class of galaxies was not well fit by the stochastic
burst models with the CF00 dust attenuation.  The black error bar in
these figures shows their mean super-colours. The models attenuated by
the Calzetti law provide a good match in all three super-colours for a
single model (2\,Gyr, 2.5$Z_\odot$, $\tau_{\rm
  V,Calzetti}$=1.2). While this does not advocate the use of the
Calzetti dust attenuation law for all blue-sequence galaxies in the
high-redshift Universe, it does show that constraints might be placed
on the shape of the attenuation law for certain classes of galaxies.
It is possible that the slightly greyer attenuation law is preferred
for the high-metallicity, old, dusty galaxies, because the high dust
column densities will lead to observed optical depths close to unity
over a wide range of wavelengths.

%------------------------------------------------------------------
\subsection{The fraction of quiescent galaxies}\label{sec:bimodal}
%------------------------------------------------------------------
The evolving fraction of quiescent galaxies with redshift is a key
constraint on galaxy evolution models. Observationally this
measurement has been approached in two ways, using rest-frame colour
selection \citep[e.g.][]{Williams:2009p8926,Ilbert:2013p9113} or
results from SED-fitting \citep[e.g.][]{Moustakas:2013p8856}. In
Section \ref{sec:LFs} and Figure \ref{fig:lfs} we presented the
super-colour version of the red-blue bimodal mass function. The blue
cloud mass function is composed of all of our star-forming classes,
including the dusty star-forming galaxies. The red sequence mass
function combines the red, post-starburst and low-metallicity classes.
From this we calculate that quiescent galaxies account for 45\% of
galaxies at high masses ($\log{\rm M^*}>11$), declining steadily to
25\% by $\log{\rm M^*}=10.5$ and 13\% at $\log{\rm M^*}=10$.

This result contrasts significantly with \citet{Ilbert:2013p9113}, who
finds a quiescent fraction of around 80\% at $\log{\rm M^*}>11$ and
similar redshifts ($0.8<z<1.1$) in the COSMOS field. They also find a
declining fraction with stellar mass, although only reaching
$\sim$20\% at $\log{\rm M^*}=10$. On the other hand, our results are
in better agreement with those of \citet{Moustakas:2013p8856} using
the PRIMUS survey who find an almost constant fraction of quiescent
galaxies, at a little over 50\%, at masses of $\log{\rm M^*}>11$. 

The difference between these results may in part be due to cosmic
variance, as well as differences in how stellar masses are
measured. The super-colour diagrams present a competitive way to
measure a robust quiescent fraction, properly accounting for dust
attenuation, as well as errors in the quantification of the SED shape
caused by incomplete sampling of the wavelength range afforded by
broad band filters.

%------------------------------------------------------------------
\subsection{Post-starburst galaxies}
%------------------------------------------------------------------

The clearest class of unusual galaxies identified in the super-colours
are the post-starburst galaxies. The excess of A and F stars in this
class gives rise to both the unique SED shape with the strong Balmer
break, and strong Balmer absorption lines that are traditionally used
to select these galaxies from spectroscopic surveys\footnote{Often
  spectral selection of post-starbursts requires them to have no
  emission lines, but this causes an unwanted bias against galaxies
  with narrow-line (obscured) AGN \citep{2006ApJ...648..281Y}, which
  are more prevalent in post-starburst galaxies than other classes
  \citep{wild_psb}, and also against galaxies where the starburst
  decays more slowly than a $\delta$ function. Selection on stellar
  continuum shape provides a more complete sample than selection on
  emission lines, but has only become possible since good
  spectrophotometric calibration has become standard in modern
  spectral surveys.}.   In Section \ref{sec:classes} we defined the
demarcation line between the post-starburst and red-sequence galaxies
by comparing spectral indices measured from available optical spectra
in the UDSz survey with a previous analysis of VVDS spectra by WWJ09
(see also Appendix \ref{app:vvds}). This ensures that our photometric
selection is similar to the WWJ09 sample, although we noted that it
may extend to slightly older ages. SED fitting does not tightly
constrain the mean burst mass fraction of the super-colour selected
post-starbursts (Figure \ref{fig:pdfs}), but the probability
distribution function is consistent with a value larger than 10\%,
similar to that measured by WWJ09.

 WWJ09 detected 16 post-starburst galaxies with $\log {\rm M^*}>9.75$
at $0.5<z<1.0$, giving a number density of
$1\times10^{-4}$Mpc$^{-3}$. In this paper, using photometric detection
in the UDS field (covering a very similar survey volume to the VVDS),
we detect 69 post-starburst galaxies with $\log{\rm M^*}>9.75$ at
$0.9<z<1.2$ and measure a slightly lower number density of
$0.45\times10^{-4}$. Assuming Poisson errors alone, we find fewer
post-starburst galaxies at $z\sim1$ than at $z\sim0.7$ at a
significance of around 2$\sigma$.  Allowing for cosmic variance and
the large completeness corrections involved in the analysis of the
spectroscopic VVDS survey, together with the uncertainty in the
relative positioning of the demarcation line separating post-starburst
galaxies from blue and red-sequence in spectroscopic and photometric
surveys, erodes the significance of this discrepancy. Overall we
conclude that, while we tentatively detect an increase in the number
of post-starburst galaxies between a mean redshift of 1 and 0.7, this
is not significant within the errors.

\citet{Whitaker:2012p8738} used the narrow-band NEWFIRM survey to
detect ``young quiescent''/post-starburst galaxies using a traditional
UVJ colour selection, finding a number density of
$3\times10^{-5}$Mpc$^{-3}$ at $z\sim1$ and factor of 10 decrease in
numbers between $z=2$ and $z=0.5$. However, the mass limit of their
sample ($\log{\rm M^*}>10.7$) is an order of magnitude higher than in the
VVDS and UDS surveys studied here. For an equivalent mass limit (using
the same IMF) we find 10 post-starburst galaxies and a number density
of $6\times10^{-6}$Mpc$^{-3}$, which is significantly lower.  However,
this is not surprising, as in Section \ref{sec:uvj} we showed that
\citet{Whitaker:2012p8738} include a large fraction of galaxies that
we class as red-sequence, based on comparison to optical
spectra.

%------------------------------------------------------------------
\subsection{Low metallicity quiescent galaxies}
%------------------------------------------------------------------

A new class of objects identified by their third super-colour are the
metal poor quiescent galaxies (Figure \ref{fig:pc123}). These are of
potential interest, because a $\sim$4\,Gyr old galaxy at $z\sim1$ has
a formation time at $z>3$. It opens the intriguing possibility that
the detection of metal-poor quiescent galaxies at $z\sim1$ could
provide useful constraints on the properties of the very first
generation of galaxies, complementary to attempts at direct detection
using their blue UV slopes ($\beta$ ) at $z>6$
\citep[e.g.][]{Bouwens:2010p6650, Dunlop:2012p8990}.

 This class of galaxies does not contribute significantly to the red
sequence luminosity or mass functions except at the very lowest masses
(Figure \ref{fig:lfs}), and they are indistinguishable from
red-sequence galaxies in traditional colour-colour
diagrams. 

Fitting BC03 stochastic burst models to the stacked SED of this class
shows that they have sub-solar metallicities and mean stellar ages of
$>3$\,Gyr. There are several possibilites for the poor fit of the BC03
stochastic burst models to this class of objects. Firstly it is
possible that the models simply do not reach low enough metallicities
(the lowest metallicity is 1/10th solar). There could be missing
components in the stellar library and/or evolutionary tracks. The M05
models have the higher SC3 values required, although the old ages of
these galaxies mean that the TP-AGB phase is unlikely to be
responsible for this. Alternatively the stochastic burst models do not
have the correct star-formation histories to describe these
populations. On the other hand, we found no significant colour
difference between Salpeter and Chabrier initial mass functions. We
also investigated the possibility that these objects were contaminants
with incorrect photometric redshifts, by removing objects with very
poorly constrained photometric redshifts. The physical parameters of
the sample did not alter significantly, and the reduced $\chi^2_\nu$
increased, indicating that incorrect redshifts are not the
cause. Ultimately, follow-up spectroscopy is needed to confirm the
nature of this class of objects.

%%%%%%%%%%%%%%%%%%%%%%%%%%%%%%%%%%%%%%%%%%%%%%%%%%%%%%%%%%%%%%%%%%
\section{Summary and Conclusions}
%%%%%%%%%%%%%%%%%%%%%%%%%%%%%%%%%%%%%%%%%%%%%%%%%%%%%%%%%%%%%%%%%%

In this paper we developed a new method to study the SED shapes of
galaxies, allowing comparison between galaxies at different redshifts
while retaining independence from spectral synthesis models. The
aim was to develop a method that allowed us to: (i) stack SEDs of
galaxies at different redshifts with the same SED shapes to better
constrain mean physical properties; (ii) compare observed colours of
galaxies without relying on the spectral synthesis models to
  be accurate, unlike traditional
K-corrections; (iii) easily visualise loss of information and biases
caused by the incomplete sampling of the rest-frame SED as a function
of redshift and SED shape; (iv) investigate the potential of broad
band photometric datasets for revealing unusual classes of galaxies. 

Application to the UDS field for $0.9<z<1.2$ galaxies results in a
very clean and tight red-sequence, and a blue-cloud that is
extended primarily by mean-stellar age and dust content. Metallicity
is orthogonal to the age-dust super-colour, meaning metallicity
can be constrained for blue-sequence galaxies for an assumed star
formation history and set of stellar population models. The reddest
end of the blue-sequence is comprised of extremely dusty galaxies that
are also metal rich and with old mean stellar ages. A strong
post-starburst population stands out clearly in SED colour space,
providing a clean method to identify galaxies which have recently
undergone a massive starburst ($\gtrsim10$\% by mass) which has
quenched, without the need for expensive spectroscopy.

We found that the BC03 spectral synthesis models provide a good
description of the SED shapes for the vast majority of $z\sim1$
galaxies, meaning that rest-frame optical-to-NIR broad band photometry is not sufficient to
distinguish more complex star formation histories and dust attenuation
laws than used here. However, we also identified some SED shapes that
are not well covered by the models: low metallicity dwarf
quiescent galaxies and massive dusty star-forming galaxies. Here we
summarise our conclusions.

\begin{itemize}
\item In modern multiwavelength broad-band photometric surveys with a
  wide wavelength coverage, the sparse sampling of galaxy SEDs caused
  by the redshifts of the galaxies does not necessarily cause a
  dramatic loss of information on their underlying SED shape. We show
  that, with currently available bands in the UDS, SED shapes can be
  recovered robustly in the redshift ranges of $0.9<z<1.2$ and
  $z>1.7$.
\item Broad-band multiwavelength photometry can detect post-starburst
  galaxies with similar properties to those detected in high-redshift
  spectroscopic surveys such as the VVDS. We detect a similar number
  density of post-starbursts more massive than $log {\rm M}^*/{\rm M}_\odot> 9.75$
  between $0.9<z<1.2$ (UDS, this paper) and $0.5<z<1$ (VVDS, WWJ09).
\item Dusty star-forming galaxies can be uniquely identified, but by
  selection they are high-metallicity ($\sim$2\,Z$_\odot$), old
  (mass-weighted mean age $\sim$2\,Gyr) and very dusty (total
  effective optical depth $\tau_V\sim$3, approximately equivalent to a
  observed optical depth to the stellar continuum of
  $\tau_V\sim$1). These galaxies have a mass function that is close to
  the red-sequence population in shape, however, we emphasise that
  this is purely a selection bias as only highest-metallicity, oldest,
  and dustiest galaxies are distinguishable to the red end of the
  blue-cloud.
\item  For the dustiest star-forming class we find better consistency
  with models attenuated by dust using the \citet{Calzetti:2000p4473}
  dust attenuation law than with the \citet{2000ApJ...539..718C}
  law. This does not support the use of the Calzetti law for all
  high-redshift star-forming galaxies, however. For extremely
  dusty galaxies a greyer dust attenuation curve may be caused by the
  high opacities to all stars at all wavelengths.
\item We tentatively identify a new class of very low-metallicity
  quiescent galaxies, which have lower luminosities and masses than
  typical red-sequence galaxies. This class is not distinguishable
  using traditional colour-colour diagrams. 
\end{itemize}

Broad band photometry contains considerable information about the
physical properties of galaxies, but simple K-corrected rest-frame
colour-colour diagrams fail to capture a significant fraction of the
information and can be biased by poor model fits. Physical properties
derived from SED fitting are prior dependent, and physical properties
will be biased when models do not fit the SED shape of the data.
Such biases are difficult to spot when errors on individual galaxies
are large, and individual galaxies are fit in isolation with no way to
group objects with similar properties prior to fitting. The method
presented here provides a complementary approach to visualise the SED
shapes of galaxy populations, allowing the stacking of populations
with the same SED shapes. We have shown it to be useful in identifying
populations with interesting physical properties, as well as
identifying areas where the spectral synthesis models need some
improvement.

\section*{acknowledgements}

UKIDSS uses the UKIRT Wide Field Camera
\citep[WFCAM][]{2007A&A...467..777C}. The photometric system is
described in \citet{2006MNRAS.367..454H}, the calibration is described
in \citet{Hodgkin:2009p4949} and the science archive is described in
\citet{2008MNRAS.384..637H}. We are indebted to the staff at UKIRT for
their tireless efforts in the face of very difficult circumstances.

V.~W. acknowledges support from the European Research Council Starting
Grant (P.I. V.~Wild), European Research Council Advanced Grant
(P.I. J.~Dunlop) and European Career Re-integration Grant
(P.I. V.~Wild). J.~S.~D. acknowledges support from the European
Research Council Advanced Grant (P.I. J. Dunlop). R.~M. acknowledges
support from the European Research Council Consolidator Grant
(P.I. R.~McLure). JSD also acknowledges the contribution of the EC FP7
SPACE project ASTRODEEP (Ref.No: 312725). This work was supported in
part by the National Science Foundation under Grant No. PHYS-1066293
and the hospitality of the Aspen Center for Physics. We would like to
thank Claudia Maraston for providing us with low metallicity stellar
population synthesis models, Paula Coelho for help identifying the
FeMg absorption feature in quiescent galaxy spectra, St\'ephane
Charlot, Tim Heckman and Henry McCracken for initial encouragement to
pursue this work, and the anonymous referee for providing a very
detailed review that helped to improve the clarity of the paper.

\footnotesize{
\bibliographystyle{mn2e}
% Use this when working: Note you have to make symbolic link to
% refs_all.bib in directory you are working
\bibliography{refs_udsz_submit_v1}

\begin{thebibliography}{}

\bibitem[\protect\citeauthoryear{Acquaviva, Gawiser \& Guaita}{Acquaviva
  et~al.}{2011}]{Acquaviva:2011p9004}
Acquaviva V.,  Gawiser E.,    Guaita L.,  2011, \apj, 737, 47

\bibitem[\protect\citeauthoryear{Allen, Hewett, Richardson, Ferland \&
  Baldwin}{Allen et~al.}{2013}]{Allen:2013p8707}
Allen J.~T.,  Hewett P.~C.,  Richardson C.~T.,  Ferland G.~J.,    Baldwin
  J.~A.,  2013, \mnras, 430, 3510

\bibitem[\protect\citeauthoryear{Berk, Shen, Yip, Schneider, Connolly, Burton,
  Jester, Hall, Szalay \& Brinkmann}{Berk et~al.}{2006}]{2006AJ....131...84V}
Berk D.~E.~V.,  Shen J.,  Yip C.-W., et~al.,  2006, \aj,
  131, 84

\bibitem[\protect\citeauthoryear{Berta, Lutz, Santini \& et al.}{Berta
  et~al.}{2013}]{Berta:2013p8847}
Berta S.,  Lutz D.,  Santini P.,    et~al., 2013, \aap, 551, 100

\bibitem[\protect\citeauthoryear{Bessell}{Bessell}{1990}]{Bessell:1990p8929}
Bessell M.~S.,  1990, \pasp, 102, 1181

\bibitem[\protect\citeauthoryear{Bolzonella, Miralles \& Pell{\'o}}{Bolzonella
  et~al.}{2000}]{Bolzonella:2000p8741}
Bolzonella M.,  Miralles J.-M.,    Pell{\'o} R.,  2000, \aap, 363, 476

\bibitem[\protect\citeauthoryear{Bouwens, Illingworth, Oesch, Trenti,
  Stiavelli, Carollo, Franx, van Dokkum, Labb{\'e} \& Magee}{Bouwens
  et~al.}{2010}]{Bouwens:2010p6650}
Bouwens R.~J.,  Illingworth G.~D.,  Oesch P.~A., et~al.
  2010, \apjl, 708, L69

\bibitem[\protect\citeauthoryear{Bowler, Dunlop, McLure, McCracken,
  Milvang-Jensen, Furusawa, Fynbo, F{\`e}vre, Holt, Ideue, Ihara, Rogers \&
  Taniguchi}{Bowler et~al.}{2012}]{Bowler:2012p9192}
Bowler R. A.~A.,  Dunlop J.~S.,  McLure R.~J., et~al.,  2012, \mnras, 426, 2772

\bibitem[\protect\citeauthoryear{Bradshaw, Almaini, Hartley, Smith, Conselice,
  Dunlop, Simpson, Chuter \& et al.}{Bradshaw
  et~al.}{2013}]{Bradshaw:2013p8772}
Bradshaw E.~J.,  Almaini O.,  Hartley W.~G., et~al., 2013, \mnras, 433, 194

\bibitem[\protect\citeauthoryear{Bruzual \& Charlot}{Bruzual \&
  Charlot}{2003}]{2003MNRAS.344.1000B}
Bruzual G.,  Charlot S.,  2003, \mnras, 344, 1000

\bibitem[\protect\citeauthoryear{Calzetti, Armus, Bohlin, Kinney, Koornneef \&
  Storchi-Bergmann}{Calzetti et~al.}{2000}]{Calzetti:2000p4473}
Calzetti D.,  Armus L.,  Bohlin R.~C.,  Kinney A.~L.,  Koornneef J.,
  Storchi-Bergmann T.,  2000, \apj, 533, 682

\bibitem[\protect\citeauthoryear{Casali, Adamson, de Oliveira \& et al.}{Casali
  et~al.}{2007}]{2007A&A...467..777C}
Casali M.,  Adamson A.,  de Oliveira C.~A.,    et~al., 2007, \aap, 467, 777

\bibitem[\protect\citeauthoryear{Charlot \& Fall}{Charlot \&
  Fall}{2000}]{2000ApJ...539..718C}
Charlot S.,  Fall S.~M.,  2000, \apj, 539, 718

\bibitem[\protect\citeauthoryear{Cirasuolo, McLure, Dunlop, Almaini, Foucaud \&
  Simpson}{Cirasuolo et~al.}{2010}]{Cirasuolo:2010p8794}
Cirasuolo M.,  McLure R.~J.,  Dunlop J.~S.,  Almaini O.,  Foucaud S.,
  Simpson C.,  2010, \mnras, 401, 1166

\bibitem[\protect\citeauthoryear{Cirasuolo, McLure, Dunlop, Almaini, Foucaud,
  Smail, Sekiguchi, Simpson \& et al.}{Cirasuolo
  et~al.}{2007}]{Cirasuolo:2007p8739}
Cirasuolo M.,  McLure R.~J.,  Dunlop J.~S., et~al., 2007, \mnras, 380, 585

\bibitem[\protect\citeauthoryear{Connolly, Genovese, Moore, Nichol, Schneider
  \& Wasserman}{Connolly et~al.}{2000}]{Connolly:2000p8848}
Connolly A.~J.,  Genovese C.,  Moore A.~W.,  Nichol R.~C.,  Schneider J.,
  Wasserman L.,  2000, arXiv:astro-ph/0008187

\bibitem[\protect\citeauthoryear{Connolly \& Szalay}{Connolly \&
  Szalay}{1999}]{1999AJ....117.2052C}
Connolly A.~J.,  Szalay A.~S.,  1999, \aj, 117, 2052

\bibitem[\protect\citeauthoryear{Connolly, Szalay, Bershady, Kinney \&
  Calzetti}{Connolly et~al.}{1995}]{1995AJ....110.1071C}
Connolly A.~J.,  Szalay A.~S.,  Bershady M.~A.,  Kinney A.~L.,    Calzetti D.,
  1995, \aj, 110, 1071

\bibitem[\protect\citeauthoryear{Conroy}{Conroy}{2013}]{Conroy:2013p8845}
Conroy, C.\ 2013, \araa, 51, 393 

\bibitem[\protect\citeauthoryear{Curtis-Lake, McLure, Pearce, Dunlop,
  Cirasuolo, Stark, Almaini, Bradshaw, Chuter, Foucaud \& Hartley}{Curtis-Lake
  et~al.}{2012}]{CurtisLake:2012p8773}
Curtis-Lake E.,  McLure R.~J.,  Pearce H.~J.,  et~al.,  2012, \mnras, 422, 1425

\bibitem[\protect\citeauthoryear{da Cunha, Charlot \& Elbaz}{da~Cunha
  et~al.}{2008}]{2008MNRAS.388.1595D}
da Cunha E.,  Charlot S.,    Elbaz D.,  2008, \mnras, 388, 1595

\bibitem[\protect\citeauthoryear{Dale, de Paz, Gordon, Hanson, Armus, Bendo,
  Bianchi, Block \& et al.}{Dale et~al.}{2007}]{Dale:2007p7234}
Dale D.~A.,  de Paz A.~G.,  Gordon K.~D.,  et~al., 2007, \apj, 655, 863

\bibitem[\protect\citeauthoryear{Dunlop, McLure, Robertson, Ellis, Stark,
  Cirasuolo \& de Ravel}{Dunlop et~al.}{2012}]{Dunlop:2012p8990}
Dunlop J.~S.,  McLure R.~J.,  Robertson B.~E.,  Ellis R.~S.,  Stark D.~P.,
  Cirasuolo M.,    de Ravel L.,  2012, \mnras, 420, 901

\bibitem[\protect\citeauthoryear{Fukugita, Ichikawa, Gunn, Doi, Shimasaku \&
  Schneider}{Fukugita et~al.}{1996}]{1996AJ....111.1748F}
Fukugita M.,  Ichikawa T.,  Gunn J.~E.,  Doi M.,  Shimasaku K.,    Schneider
  D.~P.,  1996, \aj, 111, 1748

\bibitem[\protect\citeauthoryear{Furusawa, Kosugi, Akiyama, Takata, Sekiguchi,
  Tanaka, Iwata, Kajisawa \& et al.}{Furusawa
  et~al.}{2008}]{Furusawa:2008p8792}
Furusawa H.,  Kosugi G.,  Akiyama M.,  et~al., 2008, \apjs, 176, 1

\bibitem[\protect\citeauthoryear{Gallazzi, Charlot, Brinchmann, White \&
  Tremonti}{Gallazzi et~al.}{2005}]{Gallazzi:2005p6450}
Gallazzi A.,  Charlot S.,  Brinchmann J.,  White S. D.~M.,    Tremonti C.~A.,
  2005, \mnras, 362, 41

\bibitem[\protect\citeauthoryear{Glazebrook, Offer \& Deeley}{Glazebrook
  et~al.}{1998}]{1998ApJ...492...98G}
Glazebrook K.,  Offer A.~R.,    Deeley K.,  1998, \apj, 492, 98

\bibitem[\protect\citeauthoryear{Hambly, Collins, Cross \& et al}{Hambly
  et~al.}{2008}]{2008MNRAS.384..637H}
Hambly N.~C.,  Collins R.~S.,  Cross N.~J.~G.,  et~al., 2008, \mnras, 384, 637

\bibitem[\protect\citeauthoryear{Heavens, Jimenez \& Lahav}{Heavens
  et~al.}{2000}]{2000MNRAS.317..965H}
Heavens A.~F.,  Jimenez R.,    Lahav O.,  2000, \mnras, 317, 965

\bibitem[\protect\citeauthoryear{Heckman, Kauffmann, Brinchmann, Charlot,
  Tremonti \& White}{Heckman et~al.}{2004}]{heckman04}
Heckman T.~M.,  Kauffmann G.,  Brinchmann J.,  Charlot S.,  Tremonti C.,
  White S.~D.~M.,  2004, \apj, 613, 109

\bibitem[\protect\citeauthoryear{Hewett, Warren, Leggett \& Hodgkin}{Hewett
  et~al.}{2006}]{2006MNRAS.367..454H}
Hewett P.~C.,  Warren S.~J.,  Leggett S.~K.,    Hodgkin S.~T.,  2006, \mnras,
  367, 454

\bibitem[\protect\citeauthoryear{Hodgkin, Irwin, Hewett \& Warren}{Hodgkin
  et~al.}{2009}]{Hodgkin:2009p4949}
Hodgkin S.~T.,  Irwin M.~J.,  Hewett P.~C.,    Warren S.~J.,  2009, \mnras,
  394, 675

\bibitem[\protect\citeauthoryear{Ilbert, McCracken, F{\`e}vre, Capak, Dunlop,
  Karim, Renzini, Caputi \& et al.}{Ilbert et~al.}{2013}]{Ilbert:2013p9113}
Ilbert O.,  McCracken H.~J.,  F{\`e}vre O.~L.,  et~al., 2013, \aap,
  556, 55

\bibitem[\protect\citeauthoryear{Kauffmann, Heckman, White \& et al}{Kauffmann
  et~al.}{2003}]{2003MNRAS.341...33K}
Kauffmann G.,  Heckman T.~M.,  White S.~D.~M.,  et~al., 2003, \mnras, 341, 33

\bibitem[\protect\citeauthoryear{Kewley, Groves, Kauffmann \& Heckman}{Kewley
  et~al.}{2006}]{2006MNRAS.372..961K}
Kewley L.~J.,  Groves B.,  Kauffmann G.,    Heckman T.,  2006, \mnras, 372, 961

\bibitem[\protect\citeauthoryear{Kriek, Labb{\'e}, Conroy, Whitaker, van
  Dokkum, Brammer, Franx, Illingworth, Marchesini, Muzzin, Quadri \&
  Rudnick}{Kriek et~al.}{2010}]{Kriek:2010p8971}
Kriek M.,  Labb{\'e} I.,  Conroy C.,  Whitaker K.~E., et~al., 2010, \apjl, 722, L64

\bibitem[\protect\citeauthoryear{Lane, Almaini, Foucaud, Simpson, Smail,
  McLure, Conselice, Cirasuolo \& et al.}{Lane et~al.}{2007}]{Lane:2007p8846}
Lane K.~P.,  Almaini O.,  Foucaud S.,  et~al., 2007, \mnras, 379, L25

\bibitem[\protect\citeauthoryear{Lawrence, Warren, Almaini \& et al}{Lawrence
  et~al.}{2007}]{2007MNRAS.379.1599L}
Lawrence A.,  Warren S.~J.,  Almaini O.,    et~al., 2007, \mnras, 379, 1599

\bibitem[\protect\citeauthoryear{Lu, Zhou, Wang, Wang, Dong, Zhuang \& Li}{Lu
  et~al.}{2006}]{Lu:2006p7218}
Lu H.,  Zhou H.,  Wang J.,  Wang T.,  Dong X.,  Zhuang Z.,    Li C.,  2006,
  \aj, 131, 790

\bibitem[\protect\citeauthoryear{McLure, Pearce, Dunlop, Cirasuolo,
  Curtis-Lake, Bruce, Caputi, Almaini, Bonfield, Bradshaw, Buitrago, Chuter,
  Foucaud, Hartley \& Jarvis}{McLure et~al.}{2013}]{McLure:2013p8908}
McLure R.~J.,  Pearce H.~J.,  Dunlop J.~S., et~al., 2013,
  \mnras, 428, 1088

\bibitem[\protect\citeauthoryear{Madgwick, Somerville, Lahav \& Ellis}{Madgwick
  et~al.}{2003}]{2003MNRAS.343..871M}
Madgwick D.~S.,  Somerville R.,  Lahav O.,    Ellis R.,  2003, \mnras, 343, 871

\bibitem[\protect\citeauthoryear{Maraston}{Maraston}{2005}]{Maraston:2005p8966}
Maraston C.,  2005, \mnras, 362, 799

\bibitem[\protect\citeauthoryear{Martins, Rodr{\'\i}guez-Ardila, Diniz,
  Gruenwald \& de Souza}{Martins et~al.}{2013}]{Martins:2013p9118}
Martins L.~P.,  Rodr{\'\i}guez-Ardila A.,  Diniz S.,  Gruenwald R.,    de Souza
  R.,  2013, \mnras, 431, 1823

\bibitem[\protect\citeauthoryear{Melbourne, Williams, Dalcanton, Rosenfield,
  Girardi, Marigo, Weisz, Dolphin, Boyer, Olsen, Skillman \& Seth}{Melbourne
  et~al.}{2012}]{Melbourne:2012p9128}
Melbourne J.,  Williams B.~F.,  Dalcanton J.~J., et~al.,  2012, \apj, 748, 47

\bibitem[\protect\citeauthoryear{Moustakas, Coil, Aird, Blanton, Cool,
  Eisenstein, Mendez, Wong, Zhu \& Arnouts}{Moustakas
  et~al.}{2013}]{Moustakas:2013p8856}
Moustakas J.,  Coil A.~L.,  Aird J., et~al.,  2013, \apj, 767,
  50

\bibitem[\protect\citeauthoryear{Murtagh \& Heck}{Murtagh \&
  Heck}{1987}]{1987mda..book.....M}
Murtagh F.,  Heck A.,  1987, Multivariate data analysis, ISBN 90-277-2425-3, Springer-Verlag Berlin Heidelberg

\bibitem[\protect\citeauthoryear{Noll, Burgarella, Giovannoli, Buat, Marcillac
  \& Mu{\~n}oz-Mateos}{Noll et~al.}{2009}]{Noll:2009p9084}
Noll S.,  Burgarella D.,  Giovannoli E.,  Buat V.,  Marcillac D.,
  Mu{\~n}oz-Mateos J.~C.,  2009, \aap, 507, 1793

\bibitem[\protect\citeauthoryear{Osterbrock \& Ferland}{Osterbrock \&
  Ferland}{2006}]{Osterbrock:2006p4475}
Osterbrock D.~E.,  Ferland G.~J.,  2006, Astrophysics of gaseous nebulae and
  active galactic nuclei, 2nd. ed., Sausalito CA: University Science Books

\bibitem[\protect\citeauthoryear{Pacifici, Charlot, Blaizot \&
  Brinchmann}{Pacifici et~al.}{2012}]{Pacifici:2012p9111}
Pacifici C.,  Charlot S.,  Blaizot J.,    Brinchmann J.,  2012, \mnras, 421,
  2002

\bibitem[\protect\citeauthoryear{Rudnick, Rix, Franx, Labb{\'e}, Blanton,
  Daddi, Schreiber, Moorwood, R{\"o}ttgering, Trujillo, van~der Wel, van~der
  Werf, van Dokkum \& van Starkenburg}{Rudnick
  et~al.}{2003}]{Rudnick:2003p8956}
Rudnick G.,  Rix H.-W.,  Franx M.,  Labb{\'e} I., et~al., 2003,
  \apj, 599, 847

\bibitem[\protect\citeauthoryear{Salim, Rich, Charlot, Brinchmann, Johnson,
  Schiminovich, Seibert, Mallery \& et al.}{Salim
  et~al.}{2007}]{Salim:2007p6062}
Salim S.,  Rich R.~M.,  Charlot S.,   et~al., 2007, \apjs, 173, 267

\bibitem[\protect\citeauthoryear{Schaerer \& de Barros}{Schaerer \&
  de~Barros}{2009}]{Schaerer:2009p4168}
Schaerer D.,  de Barros S.,  2009, \aap, 502, 423

\bibitem[\protect\citeauthoryear{Schmidt}{Schmidt}{1968}]{1968ApJ...151..393S} Schmidt M., 1968, \apj, 151, 
393 


\bibitem[Simpson et al.(2012)]{2012MNRAS.421.3060S} Simpson C., Rawlings 
S., Ivison R., et al., 2012, \mnras, 421, 3060 

\bibitem[\protect\citeauthoryear{Tojeiro, Heavens, Jimenez \& Panter}{Tojeiro
  et~al.}{2007}]{2007MNRAS.381.1252T}
Tojeiro R.,  Heavens A.~F.,  Jimenez R.,    Panter B.,  2007, \mnras, 381, 1252

\bibitem[\protect\citeauthoryear{Tokunaga, Simons \& Vacca}{Tokunaga
  et~al.}{2002}]{Tokunaga:2002p8948}
Tokunaga A.~T.,  Simons D.~A.,    Vacca W.~D.,  2002, \pasp, 114, 180

\bibitem[Ueda et al.(2008)]{2008ApJS..179..124U} Ueda Y., Watson M.~G., 
Stewart I.~M., et al., 2008, \apjs, 179, 124 

\bibitem[\protect\citeauthoryear{Walcher, Groves, Budav{\'a}ri \& Dale}{Walcher
  et~al.}{2011}]{Walcher:2011p8408}
Walcher J.,  Groves B.,  Budav{\'a}ri T.,    Dale D.,  2011, Astrophysics and
  Space Science, 331, 1

\bibitem[\protect\citeauthoryear{Whitaker, Kriek, van Dokkum, Bezanson,
  Brammer, Franx \& Labb{\'e}}{Whitaker et~al.}{2012}]{Whitaker:2012p8738}
Whitaker K.~E.,  Kriek M.,  van Dokkum P.~G.,  Bezanson R.,  Brammer G.,  Franx
  M.,    Labb{\'e} I.,  2012, \apj, 745, 179

\bibitem[\protect\citeauthoryear{Whitaker, Labb{\'e}, van Dokkum, Brammer,
  Kriek, Marchesini, Quadri, Franx \& et al.}{Whitaker
  et~al.}{2011}]{Whitaker:2011p8798}
Whitaker K.~E.,  Labb{\'e} I.,  van Dokkum P.~G.,  et~al., 2011, \apj, 735, 86

\bibitem[\protect\citeauthoryear{Wild, Charlot, Brinchmann, Heckman, Vince,
  Pacifici \& Chevallard}{Wild et~al.}{2011}]{Wild:2011p6890}
Wild V.,  Charlot S.,  Brinchmann J.,  Heckman T.,  Vince O.,  Pacifici C.,
  Chevallard J.,  2011, \mnras, 417, 1760

\bibitem[\protect\citeauthoryear{Wild, Heckman \& Charlot}{Wild
  et~al.}{2010}]{Wild:2010p6226}
Wild V.,  Heckman T.,    Charlot S.,  2010, \mnras, 405, 933

\bibitem[\protect\citeauthoryear{Wild \& Hewett}{Wild \&
  Hewett}{2005}]{2005MNRAS.358.1083W}
Wild V.,  Hewett P.~C.,  2005, \mnras, 358, 1083

\bibitem[\protect\citeauthoryear{Wild, Kauffmann, Heckman, Charlot, Lemson,
  Brinchmann, Reichard \& Pasquali}{Wild et~al.}{2007}]{wild_psb}
Wild V.,  Kauffmann G.,  Heckman T.,  Charlot S.,  Lemson G.,  Brinchmann J.,
  Reichard T.,    Pasquali A.,  2007, \mnras, 381, 543

\bibitem[\protect\citeauthoryear{Wild, Walcher, Johansson, Tresse, Charlot,
  Pollo, F{\`e}vre \& de Ravel}{Wild et~al.}{2009}]{Wild:2009p2609}
Wild V.,  Walcher C.~J.,  Johansson P.~H.,  Tresse L.,  Charlot S.,  Pollo A.,
  F{\`e}vre O.~L.,    de Ravel L.,  2009, \mnras, 395, 144

\bibitem[\protect\citeauthoryear{Williams, Quadri, Franx, van Dokkum \&
  Labb{\'e}}{Williams et~al.}{2009}]{Williams:2009p8926}
Williams R.~J.,  Quadri R.~F.,  Franx M.,  van Dokkum P.,    Labb{\'e} I.,
  2009, \apj, 691, 1879

\bibitem[\protect\citeauthoryear{Wong, Schawinski, Kaviraj, Masters, Nichol,
  Lintott, Keel, Darg \& et al.}{Wong et~al.}{2012}]{Wong:2012p8892}
Wong O.~I.,  Schawinski K.,  Kaviraj S.,  et~al., 2012, \mnras, 420, 1684

\bibitem[\protect\citeauthoryear{Wuyts, Labb{\'e}, Franx, Rudnick, van Dokkum,
  Fazio, Schreiber, Huang, Moorwood, Rix, R{\"o}ttgering \& van~der Werf}{Wuyts
  et~al.}{2007}]{Wuyts:2007p8937}
Wuyts S.,  Labb{\'e} I.,  Franx M., et~al.,  2007, \apj, 655, 51

\bibitem[\protect\citeauthoryear{Yan, Newman, Faber, Konidaris, Koo \&
  Davis}{Yan et~al.}{2006}]{2006ApJ...648..281Y}
Yan R.,  Newman J.~A.,  Faber S.~M.,  Konidaris N.,  Koo D.,    Davis M.,
  2006, \apj, 648, 281

\bibitem[\protect\citeauthoryear{Yip, Connolly, Berk \& et al}{Yip
  et~al.}{2004}]{2004AJ....128.2603Y}
Yip C.~W.,  Connolly A.~J.,  Berk D.~E.~V.,  et~al., 2004, \aj, 128, 2603

\bibitem[\protect\citeauthoryear{York, Adelman, Anderson \& et al}{York
  et~al.}{2000}]{2000AJ....120.1579Y}
York D.~G.,  Adelman J.,  Anderson J.,  et~al., 2000, \aj, 120, 1579

\bibitem[\protect\citeauthoryear{Zibetti, Gallazzi, Charlot, Pierini \&
  Pasquali}{Zibetti et~al.}{2013}]{Zibetti:2013p9091}
Zibetti S.,  Gallazzi A.,  Charlot S.,  Pierini D.,    Pasquali A.,  2013,
  \mnras, 428, 1479

\end{thebibliography}

}

\begin{appendix}

%------------------------------------------------------------------
\section{Post-starburst spectra in the UDSz}\label{app:vvds}
%------------------------------------------------------------------

\begin{figure*}
 \includegraphics[scale=1]{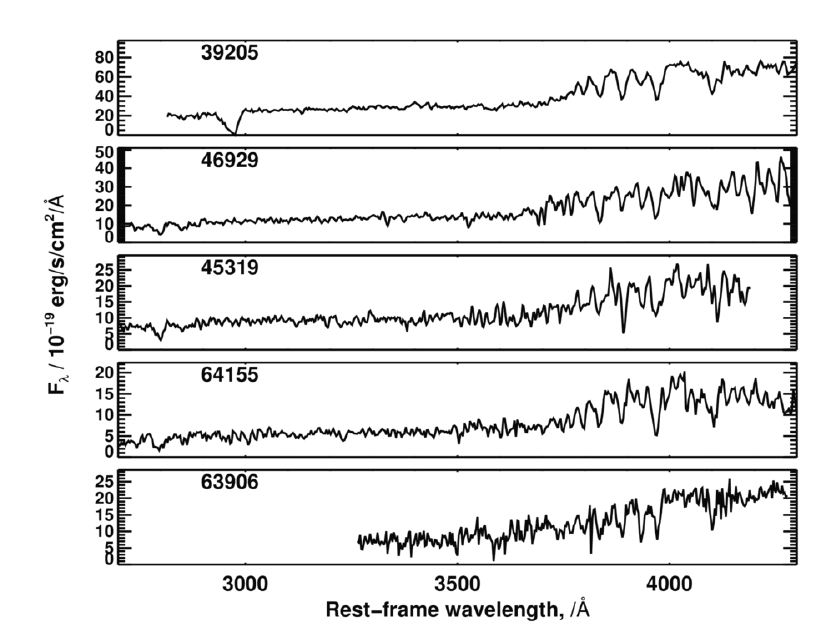}
 \includegraphics[scale=1]{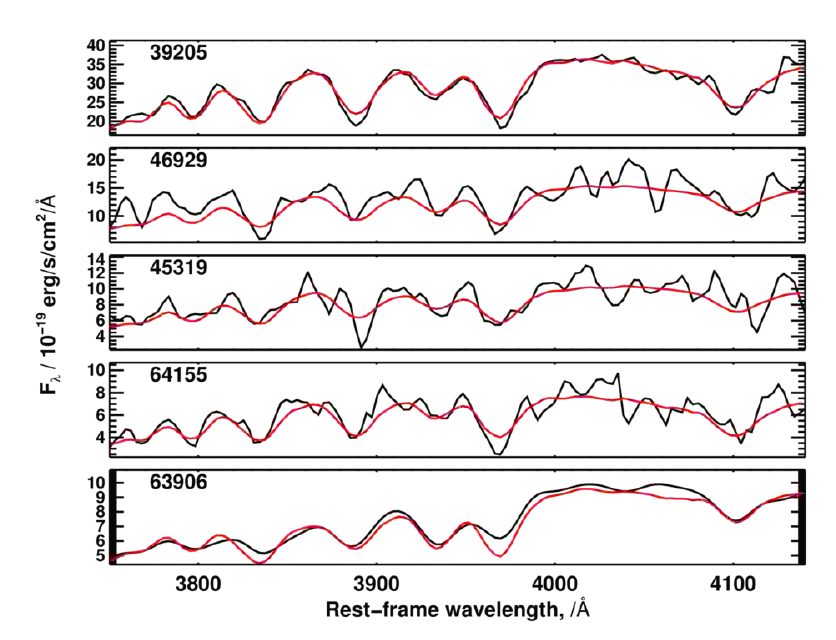}
 \caption{The five UDSz spectra of super-colour classified
   post-starburst galaxies. They all have sufficient SNR in the
   spectra to identify the strong Balmer lines and Balmer break of a
   post-starburst population. }\label{fig:vimospsbs}
 \end{figure*}

 In Figure \ref{fig:vimospsbs} we show the spectra of the five
 super-colour selected post-starburst galaxies which have UDSz
 spectra.  In the left panel the full wavelength range of the spectra
 is shown and in the right panel we focus on the region over which the
 spectral indicies are measured. The bottom spectrum has been observed
 with the FORS instrument, and the resolution has been downgraded to
 match that of the VIMOS observations. Overplotted is the
 reconstructed spectra using the first three VVDS spectral indices,
 and in Table \ref{tab:vimospsbs} are the IDs, positions, redshifts
 and VVDS spectral indices for these objects. All spectra have
 sufficient SNR that the PCA spectral indices developed in
 \citet{Wild:2009p2609} for VVDS spectra can be measured on individual
 objects as well as on the stacked spectrum, although unfortunately
 the strength of the more traditional \hda\ absorption line is not
 robustly constrained. We find that 4/5 would have been classified as
 a post-starburst galaxy by \citet{Wild:2009p2609}. The final one lies
 closer to the spectroscopically defined red-sequence, which is
 consistent with its slightly redder super-colours. 

\begin{table*}
  \caption{Details of the five super-colour selected post-starburst
    galaxies with spectra from the UDSz survey. PC1$_{\rm VVDS}$ and 
    PC2$_{\rm VVDS}$
    are the VVDS resolution spectral PCA indices calculated using the
    eigenvectors and method presented for galaxies in the VVDS survey by
    Wild et al. (2009). }\label{tab:vimospsbs}
\begin{tabular}{cccccccccc}
UDSz-ID  & RA & Dec & $z_{phot}$ &$ z_{spec}$ & SC1 & SC2 & SC3 & PC1$_{\rm VVDS}$ &
PC2$_{\rm VVDS}$ \\\hline
         39205&       34.536512&      -5.1441217&0.98&1.01&-3.5$\pm$0&4.9$\pm$0&-0.1$\pm$0&-0.11$\pm$0.04&0.85$\pm$0.04\\
       46929&       34.538982&      -5.0721098&1.05&1.06&-1.7$\pm$0&6.9$\pm$0&0.3$\pm$0&0.17$\pm$0.08&0.74$\pm$0.09\\
       45319&       34.308255&      -5.0884575&1.17&1.15&0.1$\pm$0&5.1$\pm$0&0.2$\pm$0&-0.02$\pm$0.10&0.83$\pm$0.09\\
       64155&       34.409449&      -4.9054747&1.15&1.17&-2.5$\pm$0&7.4$\pm$0&0.7$\pm$0&-0.17$\pm$0.1&1.3$\pm$0.1\\
       63906&       34.704012&      -4.9078703&1.16&1.17&-11.0$\pm$0&4.2$\pm$0&0.9$\pm$0&0.87$\pm$0.07&0.37$\pm$0.06\\

\end{tabular}
 \end{table*}

%------------------------------------------------------------------
\section{The K-band luminosity function}\label{app:KLF}
%------------------------------------------------------------------
In the main paper we present rest-frame 1\mum\ luminosities, which are
within the observed wavelength range of the data, and therefore do not
require extrapolation using spectral synthesis models. To enable
direct comparison with previous work we provide here the rest-frame
K-band luminosity function of each class, with K-band luminosity of
each galaxy extrapolated using the best-fit BC03 stochastic burst
spectral synthesis models to the first three super-colours of each
galaxy.

\begin{figure}
\includegraphics[scale=1]{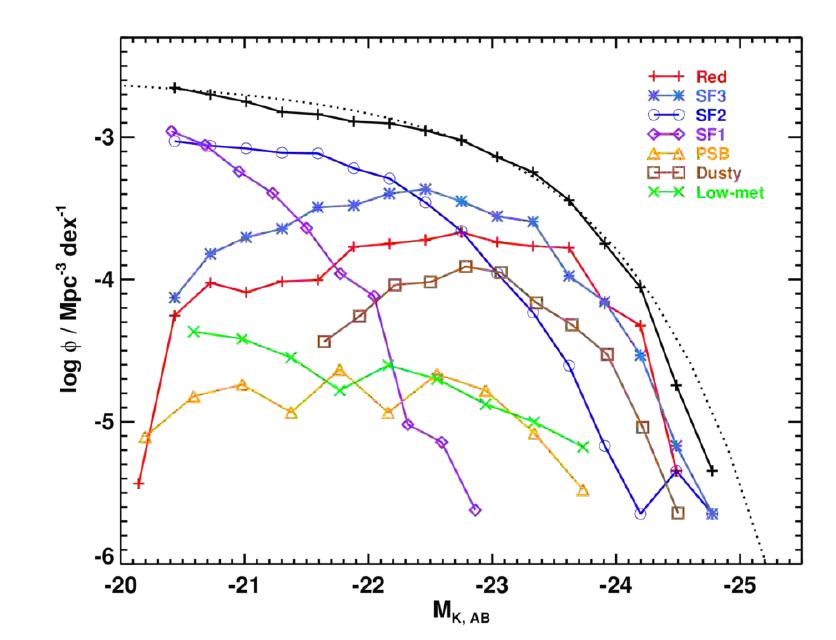}
\caption{The K-band luminosity functions of the complete sample
  (black) and individual classes as labelled (colours). The dotted
  line is the fit to the measured luminosity function at $z=1$ from
  \citet{Cirasuolo:2010p8794} for an earlier data release of the same
  field. Overall agreement is good, with differences consistent within
  the errors. }
\end{figure}

\end{appendix}

\end{document}